\documentstyle[editedvolume,numreferences,epsfig]{crckapb}

\newcommand{\bm}[1]{\mbox{\boldmath $#1$}}
\newcommand{\bt}[1]{{\bm #1}_{_T}}
\newcommand{\Pslash}{\kern 0.2 em P\kern -0.56em \raisebox{0.3ex}{/}}
\newcommand{\pslash}{\kern 0.2 em p\kern -0.4em /}
\newcommand{\nslash}{\kern 0.2 em n\kern -0.4em /}
\newcommand{\kslash}{\kern 0.2 em k\kern -0.45em /}
\newcommand{\Sslash}{\kern 0.2 em S\kern -0.56em \raisebox{0.3ex}{/}}
\newcommand{\dslash}{\kern 0.2 em \partial\kern -0.56em \raisebox{0.3ex}{/}}
\newcommand{\xbj}{x_{_B}}
\newcommand{\zh}{z_h}
\def\be{\begin{equation}}
\def\ee{\end{equation}}
\def\bea{\begin{eqnarray}}
\def\ba{\begin{eqnarray}}
\def\eea{\end{eqnarray}}
\def\ea{\end{eqnarray}}
\def\st{{\scriptscriptstyle T}}
\def\slash{\rlap{/}}

\begin{opening}
\title{STRUCTURE OF HADRONS IN HARD PROCESSES}

\author{P.J. MULDERS}
\institute{Division of Physics and Astronomy\\
Faculty of Sciences, Vrije Universiteit\\
De Boelelaan 1081, 1081 HV Amsterdam, Netherlands}

\end{opening}

\runningtitle{STRUCTURE OF HADRONS IN HARD PROCESSES}

\begin{document}


\begin{abstract}
In these lectures I want to discuss how the structure functions in
deep inelastic scattering relate to quark and gluon correlation functions.
In particular we will consider the issue of intrinsic transverse
momenta of quarks, which becomes important in processes like
1-particle inclusive leptoproduction. Some examples of cross
sections and asymmetries, in particular in polarized scattering
processes are discussed.
\end{abstract}

\section{Introduction}

The central point of these lectures is the availability of a field
theoretical
framework for the strong interactions. It is the nonabelian gauge theory
based on the color symmetry group $SU(N_c)$ with $N_c$ = 3. This theory,
Quantum Chromodynamics (QCD) underlies the strong interactions
of which the existence is known since the thirties. At that time the goal
was to understand the forces in the atomic nucleus. The experimental
results of many experiments revealed the existence of a rich spectrum of
hadrons, baryons and mesons.
The presence of these excitation spectra and also explicit
measurements of the charge and current distributions of the hadrons revealed
the composite nature of hadrons.
The quark model brought some order into this situation, describing baryons as
composite systems of three valence quarks and mesons as a composite system
of a valence quark and antiquark. Valence refers here to the fact that
these quarks are the contributors to the quantum numbers (upness,
downness, strangeness, etc.) of the hadrons. The symmetry considerations
based on flavor $SU(3)$ and flavor-spin $SU(6)$ are the basis of the
success of the quark model. Color was introduced in a rather early stage
to solve a number of problems such as the fermion statistics of the quarks.
The experimental results of the SLAC-MIT deep inelastic electron scattering
experiments, which indicated the existence of hard 'point-like' constituents
in the nucleon and the field-theoretical developments
in nonabelian gauge theories, specifically the notion of asymptotic freedom
and the proof of renormalizability, led to the natural emergence of
QCD as the theory for the strong interactions between the quarks and gluons.
The important feature of QCD is the fact that the force becomes weaker at
short distances. This anti-screening behavior or asymptotic freedom is a
unique feature of non-abelian gauge theories.

QCD is part of the Standard Model that describes the strong and electroweak
forces. All forces in the Standard Model are described within the framework
of nonabelian gauge theories, based on a gauge symmetry.
For testing directly QCD, one preferably avoids the presence of
hadrons, as these constitute complex bound state systems. Useful scattering
processes for this purpose are
\begin{tabbing}
aaaaaaaaaaaaaaaaaaaaaaaaaa\=bbbbbbbb \kill
$e^+ e^- \longrightarrow X$ \>(inclusive annihilation), \\
$e^+ e^- \longrightarrow$ jets \>(jet production), \\
$\ell H \longrightarrow \ell^\prime X$ \>(inclusive leptoproduction).
\end{tabbing}
Analyzing jets produced in $e^+e^-$ scattering represent, modulo
complications arising from hadronization, hard scattering processes
$e^+e^-$ $\rightarrow$
$q \bar q$, $e^+e^-$ $\rightarrow$ $q \bar q g$, etc. For these processes
perturbative QCD can be used to do reliable calculations. Similarly one can
use perturbative QCD to study the $\ln Q^2$ dependence of the elementary
$\ell q$ $\rightarrow$ $\ell^\prime q$ cross sections that are important
in leptoproduction.

The properties of hadrons are poorly described directly starting from
QCD. They require a nonperturbative approach in quantum field theory.
This has led to the use of models that incorporate some features like
confinement and asymptotic freedom and symmetries of the underlying theory,
such as quark potential models, bag models or chiral models.
Most promising from a fundamental point of view are lattice gauge
approaches.

Controlling and selectively probing the nonperturbative regime in high
energy scattering processes is the key to study the structure of hadrons
in the context of QCD. The control parameters for the target and the probe
are the spin and flavor, which in combination with
the kinematical flexibility in scattering processes is used to select the
observable and its gluonic or quarkic nature. Examples are
\begin{tabbing}
aaaaaaaaaaaaaaaa
\= bbbbbbbbbbbbbbbbbbbbbb
\= cccccccccccc \kill
$\ell H  \longrightarrow  \ell^\prime H$
\>(elastic leptoproduction)
\>(spacelike) form factors, \\
$\ell H  \longrightarrow  \ell^\prime X$
\>(inclusive leptoproduction)
\>distribution functions, \\
$\ell H \longrightarrow \ell^\prime h X$
\>(1-particle inclusive
\>distribution and \\
\>\quad leptoproduction)
\> \quad fragmentation functions, \\
$e^+ e^- \longrightarrow h \bar h$
\>(annihilation into $h\bar h$)
\>(timelike) form factors, \\
$e^+ e^- \longrightarrow h X$
\>(1-particle inclusive
\>fragmentation functions, \\
\>\quad annihilation)
\> \\
$H_1 H_2 \longrightarrow \mu^+ \mu^- X$
\>(Drell-Yan scattering)
\>distribution functions.
\end{tabbing}

These notes consist of two parts. After this introduction we discuss
the basic formalism to deal with the beforementioned processes, introducing
the hadron tensor, form factors, structure functions. These are needed
to write down general expressions for the electroweak cross sections.
The second part reviews the introduction of parton distribution and
fragmentation functions that enable a systematic treatment of the
structure functions at high energies and/or momentum transfer.

\section{Inclusive Leptoproduction}

\subsection{The hadron tensor}

\begin{figure}[b]
\begin{center}
\begin{minipage}{6 cm}
\epsfig{file=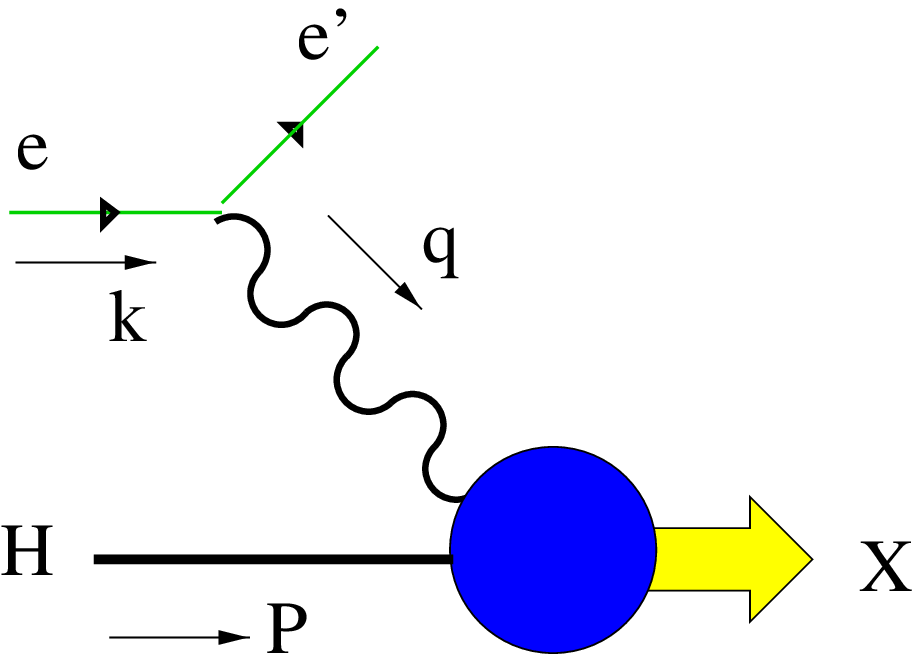,width = 4.5 cm}
\end{minipage}
\begin{minipage}{3 cm}
\begin{eqnarray*}
&&\ell H \longrightarrow \ell^\prime X
\end{eqnarray*}
\begin{eqnarray*}
&&\xbj = \frac{Q^2}{2P\cdot q} \\
&&y = \frac{P\cdot q}{P\cdot k}
\end{eqnarray*}
\end{minipage}
\end{center}
\caption{\label{fig0}
Momenta and invariants in inclusive leptoproduction. The scale is
set by the invariant momentum squared of the virtual photon, $q^2
\equiv -Q^2$, which for a hard process becomes $Q^2 \rightarrow \infty$.}
\end{figure}

For the process $\ell + H \rightarrow \ell^\prime + X$
(see Fig.~\ref{fig0}), the cross section can be separated into a lepton and
hadron part. Although the lepton part is simpler, let us start
with the hadron part, 
\bea
&&2M\,W_{\mu\nu}^{(\ell H)}( q; {P S} )
=\frac{1}{2\pi}
\sum_X \int \frac{d^3 P_X}{(2\pi)^3 2P_X^0}
(2\pi)^4 \delta^4 (q + P - P_X)
\nonumber
\\
&& \hspace{5 cm} \times
\langle {P S} |{J_\mu (0)}|P_X \rangle
\langle P_X |{J_\nu (0)}|{P S} \rangle,
\eea
The simplest thing to do is to parametrize this tensor in standard 
tensors and structure functions. Instead of the traditional 
choice~\cite{Roberts} for
these tensors, $g_{\mu\nu}$ and $P_\mu P_\nu$ and structure functions 
$W_1$ and $W_2$, we immediately go to a dimensionless representation, 
using a natural space-like momentum (defined by $q$) and a time-like 
momentum constructed from $P$ and $q$,
\bea
\hat z^\mu = -\hat q^\mu &=& -\frac{q^\mu}{Q},
\\
\hat t^\mu &=& \frac{\tilde P^\mu}{\sqrt{\tilde P^2}} =
\frac{1}{\sqrt{\tilde P^2}}
\,\left(P^\mu - \frac{P\cdot q}{q^2}\,q^\mu\right)
= \frac{q^\mu + 2\xbj\,P^\mu}{Q}.
\eea
Using hermiticity for the currents, parity invariance and current 
conservation one obtains as the most general form the symmetric 
tensor 
\bea
M\,W_S^{\mu\nu}({q,P}) =
\underbrace{\left\lgroup - g^{\mu\nu} +\hat q^\mu \hat q^\nu 
-\hat t^\mu \hat t^\nu\right\rgroup}_{-g_\perp^{\mu\nu}}{F_1}
+ \hat t^\mu\hat t^\nu
\,\underbrace{\left(\frac{F_2}{2\xbj}-{F_1}\right)}_{F_L} ,
\label{param}
\eea
where the structure functions $F_1$ and $F_2$ or the transverse and 
longitudinal structure functions, $F_T = F_1$ and $F_L$, depend only 
on the for the hadron part relevant invariants $Q^2$ and $\xbj$.
In all equations given here we have omitted target mass effects of 
order $M^2/Q^2$.

\subsection{The lepton tensor}

In order to write down the cross section one needs to include the 
necessary phase space factors and include the lepton part given by 
the tensor 
\bea
L_{\mu\nu}^{(\ell H)} (k \lambda ; k^\prime \lambda^\prime)
= 2 k_{\mu}k^\prime_{\nu} + 2 k_{\nu}k^\prime_{\mu} 
- Q^2 g_{\mu\nu} +2i\lambda_e\, \epsilon_{\mu\nu\rho\sigma}
q^\rho k^{\sigma} .
\eea
We have included here the (longitudinal) lepton polarization 
($\lambda_e = \pm 1$). For later convenience it is useful to rewrite 
this tensor also in terms of the space-like and time-like vectors 
$\hat q$ and $\hat t$. It is a straightforward exercise to get
\bea
k^\mu = \frac{Q}{2}\,\hat q^\mu + \frac{(2-y) Q}{2y}\,\hat t^\mu
+ \frac{Q\sqrt{1-y}}{y}\,\hat \ell^\mu, 
\eea
where $\hat \ell$ is the perpendicular direction defining the lepton 
scattering plane (see Fig.~\ref{fig1}). This perpendicular direction 
becomes relevant only if other vectors than $P$ and $q$ are present, e.g.
a spin direction in polarized 
scattering or the momentum of a produced hadron in 1-particle 
inclusive processes. The lepton tensor becomes
\bea
L^{\mu \nu}_{(\ell H)} & = &\frac{Q^2}{y^2} \Biggl[ 
-2 \left( 1 - y + \frac{1}{2}\,y^2 \right) g_\perp^{\mu \nu}
+ 4(1-y) \hat t^\mu \hat t^\nu
\nonumber \\ && \qquad 
+ 4(1-y)\left( \hat \ell^\mu\hat \ell^\nu 
+\frac{1}{2}\,g_\perp^{\mu \nu}\right)
+ 2(2 - y)\sqrt{1-y}\,\,\hat t^{\{ \mu}\hat \ell^{\nu \}}
\nonumber \\ &&\qquad 
-i\lambda_e\,y(2-y)\,\epsilon_\perp^{\mu \nu}
- 2i\lambda_e\,y\sqrt{1-y}\,\,\hat \ell_\rho \epsilon_\perp^{\rho\,[ \mu}
\hat t^{\nu ]} \Biggr],
\label{leptontensor}
\eea
where $\epsilon_\perp^{\mu\nu} \equiv 
\epsilon^{\mu\nu\rho\sigma}\hat t_\rho \hat q_\sigma$.

\subsection{The inclusive cross section}

The cross section for unpolarized lepton and hadron only involves 
the first two (symmetric) terms in the lepton tensor and one obtains
\bea
\frac{d\sigma_{OO}}{d\xbj dy} & = & \frac{4\pi\,\alpha^2\,\xbj s}{Q^4}
\Biggl\{ \left( 1-y + \frac{1}{2}\,y^2\right) {F_T(\xbj,Q^2)}
\nonumber \\
&& \mbox{} \hspace{2cm}
+ \left( 1 -y\right) {F_L(\xbj,Q^2)}\Biggr\}.
\eea
As soon as the exchange of a $Z^0$ boson becomes important the 
hadron tensor is no longer constrained by parity invariance and a 
third structure function $F_3$ becomes important. 

\subsection{Target polarization}

The use of polarization in leptoproduction provides new ways to probe 
the hadron target. 
For a spin 1/2 particle the initial state is described by a
2-dimensional spin density matrix $\rho = \sum_\alpha \vert \alpha\rangle
p_\alpha \langle \alpha\vert$ describing the probabilities $p_\alpha$
for a variety of spin possibilities. This density matrix is hermitean 
with Tr$\,\rho$ = 1. It can in the target rest frame be expanded
\be
\rho_{ss^\prime} = \frac{1}{2}
\left( 1 + \bm S\cdot \bm \sigma_{ss^\prime}\right),
\ee
where $\bm S$ is the spin vector. When 
$\vert \bm S\vert = 1$ one has a pure state (only one state $\vert
\alpha\rangle$ and $\rho^2 = \rho$), when $\vert \bm S\vert \le 1$ 
one has an ensemble of states. For the
case $\vert \bm S\vert = 0$ one has simply an averaging over spins,
corresponding to an unpolarized ensemble. To include spin one
could generalize the hadron tensor to a matrix in spin space,
$\tilde W_{s^\prime s}^{\mu\nu}(q,P)$ $\propto$
$\langle P, s^\prime\vert J^\mu\vert X\rangle\langle X\vert 
J^\nu\vert P, s>$
depending only on the momenta or one can look at the tensor
$\sum_\alpha p_\alpha \tilde W_{\alpha\alpha}^{\mu\nu}(q,P)$,
which is given by
\be
W^{\mu\nu}(q,P,S) = \mbox{Tr}\left( \rho(P,S) \tilde W^{\mu\nu}(q,P)
\right),
\label{defS}
\ee
with the spacelike spin vector $S$ appearing {\em linearly} and
in an arbitrary frame satisfying $P\cdot S = 0$. It has
invariant length $-1 \le S^2 \le 0$.
It is convenient to write the spin vector as
\bea
S^\mu = \frac{\lambda}{M}\left( P^\mu - \frac{2\xbj 
M^2}{Q^2}\,q^\mu\right) + S_\perp^\mu,
\label{inclspin}
\eea
with $\lambda = M(S\cdot q)/(P\cdot q)$. For a pure state one
has $\lambda^2 + \bm S_\perp^2 = 1$.
Using symmetry constraints one obtains 
for electromagnetic interactions (parity conservation)
an antisymmetric part in the hadron tensor, 
\bea
M\,W_A^{\mu\nu}({q,P,S}) =
\underbrace{-i\,{\lambda}\,
\frac{\epsilon^{\mu\nu\rho\sigma}P_\rho q_\sigma}
{P\cdot q}}_{-i\,{\lambda}\,\epsilon_\perp^{\mu\nu}}
\,{g_1}
+ i\,\frac{2M\xbj}{Q}
\,\hat t_{\mbox{}}^{\,[\mu}\epsilon_\perp^{\nu ]\rho} {S_{\perp\rho}}
\,{g_T}.
\label{wanti}
\eea
The polarized part of the cross section becomes
\bea
\frac{d\sigma_{LL}}{d\xbj dy} & = & \lambda_e\,\frac{4\pi\,\alpha^2}{Q^2}
\Biggl\{ {\lambda}\,\left( 1-\frac{y}{2} \right){g_1(\xbj,Q^2)}\nonumber
\\ && \qquad\qquad
-\vert {S_\perp}\vert\,\cos\,\phi_S^\ell\,\frac{2M\xbj}{Q} \sqrt{1-y}
\,\,{g_T(\xbj,Q^2)}\Biggr\}.
\eea

\section{Semi-inclusive leptoproduction}

\subsection{The hadron tensor}

\begin{figure}[t]
\begin{center}
\begin{minipage}{8.5cm}
\epsfig{file=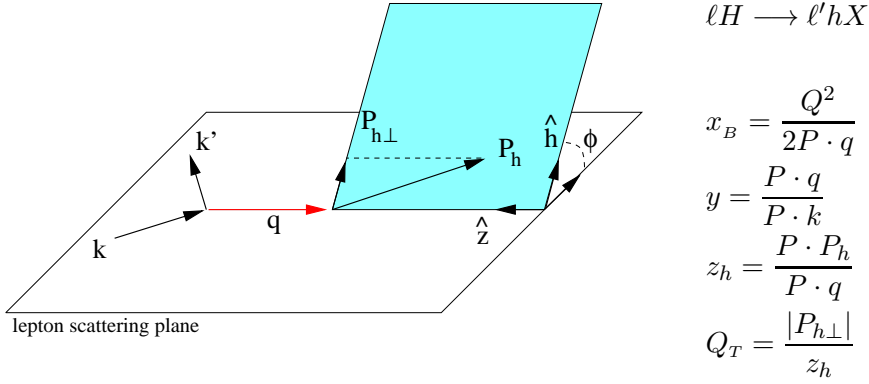,width=8.5cm}
\end{minipage}
\begin{minipage}{2.5 cm}
\begin{eqnarray*}
&&\ell H \longrightarrow \ell^\prime h X
\end{eqnarray*}
\begin{eqnarray*}
&&\xbj = \frac{Q^2}{2P\cdot q} \\
&&y = \frac{P\cdot q}{P\cdot k}\\
&&z_h = \frac{P\cdot P_h}{P\cdot q}\\
&&Q_\st = \frac{\vert P_{h\perp}\vert}{z_h}
\end{eqnarray*}
\end{minipage}
\end{center}
\caption{\label{fig1}
Kinematics for 1-particle inclusive leptoproduction.}
\end{figure}

More flexibility in probing
new aspects of hadron structure is achieved in semi-inclusive 
scattering processes. For instance in 1-particle inclusive 
measurements one can measure azimuthal dependences in the cross 
sections. The central
object of interest for 1-particle
inclusive leptoproduction, the hadron tensor, is given by
\bea
&&2M{\cal W}_{\mu\nu}^{(\ell H)}( q; {P S; P_h S_h} )
=
\sum_X \int \frac{d^3 P_X}{(2\pi)^3 2P_X^0}
\delta^4 (q + P - P_X - P_h)
\nonumber
\\
&& \hspace{2.5 cm} \times
\langle {P S} |{J_\mu (0)}|P_X; {P_h S_h} \rangle
\langle P_X; {P_h S_h} |{J_\nu (0)}|{P S} \rangle,
\eea
where $P,\ S$ and $P_h,\ S_h$ are the momenta and spin vectors
of target hadron and produced hadron,
$q$ is the (space-like) momentum transfer with $-q^2$ = $Q^2$ sufficiently
large.
The kinematics is illustrated in Fig.~\ref{fig1}, where also the scaling
variables are introduced.
For the parametrization of the hadron tensor in terms of structure 
functions it is useful to introduce the directions $\hat q$ and 
$\hat t$ as before and using the vector $P_h$ to construct
a vector that is orthogonal to these vectors. For the situation that
$P\cdot P_h$ is ${\cal O}(Q^2)$ (current fragmentation!) one
finds that
\bea
q_\st^\mu = q^\mu +\xbj\,P^\mu - \frac{P_{h}^\mu}{\zh} =
-\frac{P_{h\perp}^\mu}{\zh} \equiv -Q_T\,\hat h^\mu ,
\eea
is such a vector.
This vector is proportional to the transverse momentum of the
outgoing hadron with respect ot $P$ and $q$. It can also be 
considered as
the transverse momentum of the photon with respect to the hadron
momenta $P$ and $P_h$.
For an unpolarized (or spin 0) hadron in the final state
the symmetric part of the tensor is given by
\bea
M{\cal W}_S^{\mu\nu}({q,P,P_h}) &=&
- g_\perp^{\mu\nu}\,{\cal H}_T
+ \hat t^\mu\hat t^\nu\,{\cal H}_L
\nonumber \\ &&
+ \hat t^{\,\{\mu}\hat h^{\nu\}}\,{\cal H}_{LT}
+ \left\lgroup 2\,\hat h^\mu \hat h^\nu + g_\perp^{\mu\nu}\right\rgroup
{\cal H}_{TT} .
\eea
Noteworthy is that also an antisymmetric term in the tensor is allowed,
\bea
M{\cal W}_A^{\mu\nu}({q,P,P_h}) =
- i\hat t^{\,[\mu}\hat h^{\nu]}\,{\cal H}^\prime_{LT}.
\label{sidiswanti}
\eea

\subsection{The semi-inclusive cross section}

Clearly the lepton tensor in Eq.~\ref{leptontensor} is able to distinguish
all the structures in the semi-inclusive hadron tensor. The 
symmetric part gives the cross section for unpolarized leptons,
\bea
\frac{d\sigma_{OO}}{d\xbj dy\,d\zh d^2q_\st}
& = & \frac{4\pi\,\alpha^2\,s}{Q^4}\,\xbj \zh
\Biggl\{ \left( 1-y+\frac{1}{2}\,y^2 \right){\cal H}_T
+ (1-y)\,{\cal H}_L
\nonumber\\ && \qquad \qquad \qquad
\mbox{} - (2-y)\sqrt{1-y}\,\cos \phi_h^\ell\,\,{\cal H}_{LT}
\nonumber\\ && \qquad \qquad \qquad
\mbox{} + (1-y)\,\cos 2\phi_h^\ell\,\,{\cal H}_{TT}
\Biggr\}
\eea
while the antisymmetric part gives
the cross section for a polarized lepton 
(note the target is not polarized!) 
\bea
\frac{d\sigma_{LO}}{d\xbj dy\,d\zh d^2q_\st}
= \lambda_e\,\frac{4\pi\,\alpha^2}{Q^2}\,\zh
\,\sqrt{1-y}\,\sin \phi_h^\ell\,\,{\cal H}^\prime_{LT} .
\eea
Of course many more structure functions appear for polarized targets or
if one considers polarimetry in the final state.

\section{Form factors}

A special case is the situation in which the final state is identical
to the initial state, elastic scattering. In that case the final state
four momentum is $P^\prime = P+q$ and is fixed to be $(P+q)^2 = M^2$,
i.e. $\xbj = 1$. We can still use the formalism for inclusive leptoproduction
but the hadron tensor becomes
becomes
\be
2M\,W_{\mu\nu}(q,P) = 
\underbrace{\langle P |{J_\mu (0)}|P^\prime \rangle
\langle P^\prime |{J_\nu (0)}|P \rangle}_{H_{\mu\nu}(P;P^\prime)}
\,\frac{1}{Q^2}\,\delta (1-\xbj).
\ee
One needs the current matrix elements of the electromagnetic current,
which using hermiticity, parity and current conservation 
can be parametrized in terms of (real) form factors,
for a spin 1/2 particle
\be
\langle P^\prime, S^\prime \vert J_\mu (x) \vert P,S \rangle =
e^{i\,q\cdot x}\,\bar U_{S^\prime} (P^\prime) 
\underbrace{\left[
\gamma_\mu\, F_1(Q^2) 
+ \frac{i\sigma_{\mu\nu}q^\nu}{2M}\, F_2(Q^2) 
\right]}_{\Gamma_\mu(P,P^\prime)}
\,U_S(P), 
\ee
where $U_S(P)$ are the standard Dirac spinors.

In order to interpret the form factors for spacelike $q$,
note that in leptoproduction there always exist a frame in which $q$ 
is purely spacelike ($q^0$ = 0), the socalled brick-wall or Breit frame.
Working out the current expression for the nucleon in this frame
($P^{\prime 0} = P^0$, $\bm P = -\bm q/2$, $\bm P^\prime = +\bm q/2$), 
gives
\ba
\langle P^\prime, S^\prime \vert J_0^{em} \vert P,S \rangle & = &
2M\,\left[ F_1(Q^2) - \frac{Q^2}{4M^2}\,F_2(Q^2)\right] e^{iq\cdot x}
\nonumber \\ & \equiv &
2M\, G_E(Q^2) e^{iq\cdot x}, \\
\langle P^\prime, S^\prime \vert {\bm J}^{em} \vert P,S \rangle & = &
\left[ F_1(Q^2) + F_2(Q^2) \right] 
\,(i{\bm \sigma}_N\, \times \bm q) e^{iq\cdot x}
\nonumber \\
& \equiv & 
G_M(Q^2)\,(i{\bm \sigma}_N \times \bm q) e^{iq\cdot x},
\ea
where $\chi_{S^\prime}^\dagger\,{\bm \sigma}\,\chi_S$ $\equiv$
${\bm \sigma}_N$. These expressions show the relevance of the Sachs
form factors $G_E$ and $G_M$ being the Fourier transfer of the
spatial charge and current distribution. 
The quantity $e\,G_E(0)$ is the charge of the nucleon, $e\,G_M(0)/M$ 
is the  magnetic moment of the nucleon. The quantity $\kappa$ = $F_2(0)$ 
= $G_M(0)-G_E(0)$ is the anomalous magnetic moment.
\begin{quotation}
\small
One has for a point-particle (e.g. electron or muon) 
(in lowest order in $\alpha$)
\[
F_1(0) = G_E(0) = 1 , \qquad F_2(0) = 0, \qquad G_M(0) = 1,
\]
and no $Q^2$-dependence,
while for a composite particle like the nucleon one has a combination of
quark form factors weighted by the charges, giving
\begin{eqnarray*}
&& F_1^p(0) = G_E^p(0) = 1, \qquad F_2^p(0) = \kappa_p \approx 1.79,
\quad \qquad G_M^p = \mu_p \approx 2.79,\\
&& F_1^n(0) = G_E^n(0) = 0, \qquad F_2^n(0) = \kappa_n \approx -1.91,
\qquad G_M^n = \mu_n \approx -1.91.
\end{eqnarray*}
The $Q^2$-dependence for the electromagnetic form factors of the nucleon
approximately is given by
\[
G_E^p(Q^2) \approx \frac{G_M^p(Q^2)}{\mu_p} \approx
\frac{G_M^n(Q^2)}{\mu_n}
\approx \frac{1}{\left( 1 + Q^2/0.69\ \mbox{GeV}^2\right)^2}
\]
(dipole form factors).
\end{quotation}
For the tensor $H_{\mu \nu}$ one obtains the result
\begin{eqnarray}
H_{\mu \nu}(P,P^\prime) & = & \frac{1}{2} \sum_{S,S^\prime}
\langle P,S \vert J_\mu (0) \vert P^\prime,S^\prime \rangle \langle
P^\prime,S^\prime \vert J_\nu(0) \vert P,S \rangle \nonumber \\
& = & \frac{1}{2}\,Tr\ \left[\Gamma_\mu (\slash P^\prime + M) \Gamma_\nu
(\slash P + M) \right] \nonumber \\
& = & \left( \frac{q_\mu q_\nu}{q^2} - g_{\mu \nu} \right)
\,Q^2 (F_1 + F_2)^2
+ 4\,\tilde P_\mu \tilde P_\nu\,\left( F_1^2 
+ \frac{Q^2}{4M^2}\,F_2^2 \right)
\nonumber \\ & = &  - g_{\perp\,\mu \nu}\,Q^2
G_M^2 + \hat t_\mu \hat t_\nu\,4M^2\,G_E^2
\end{eqnarray}
One thus sees the following {\em elastic} contribution in
the structure functions 
\bea
&&F_T(\xbj,Q^2) = G_M^2(Q^2)\,\delta(1-\xbj),
\\
&&F_L(\xbj,Q^2) = \frac{4M^2}{Q^2}\,G_E^2(Q^2)\,\delta(1-\xbj).
\eea

In particular when one is also considering other currents than the 
electromagnetic case, it is useful to realize that the currents for
the $\gamma$, $Z$- or $W$-particles of course all are known in terms
of quark vector and axial vector currents, e.g. 
\bea
&&J_\mu^{(\gamma)}(x) = \sum_{q}Q\,V_\mu^{(q)}(x)
\\
&&J_\mu^{(Z)}(x) = \sum_{q}\left(I_W^3 - 2Q\,\sin^2\theta_W \right)
\,V_\mu^{(q)}(x) - I_W^3\,A_\mu^{(q)}(x),
\\
&&J_\mu^{(W^\pm)}(x) = \sum_{q}
I_W^\pm\left(V_\mu^{(q)}(x) - A_\mu^{(q)}(x)\right),
\eea
where the vector and axial vector quark currents are
\bea
&&V_\mu^{(q)}(x) = \overline \psi(x)\,\gamma_\mu \psi(x),
\\
&&A_\mu^{(q)}(x) = \overline \psi(x)\,\gamma_\mu\gamma_5 \psi(x),
\eea
One can of course also consider a parametrization of these currents
in terms of form factors for a particular flavor, again using
hermiticity, parity and the conservation of the vector current.
For a spin 1/2 particle (e.g. the nucleon) one obtains
\begin{eqnarray}
&&\langle P^\prime, S^\prime \vert V_\mu (0) \vert P,S \rangle =
\bar U_{S^\prime} (P^\prime) 
\underbrace{\left[
\gamma_\mu\, F^q_1(Q^2) 
+ \frac{i\sigma_{\mu\nu}q^\nu}{2M}\, F^q_2(Q^2) 
\right]}_{\Gamma_\mu^V(P,P^\prime)}
\,U_S(P), \nonumber \\
&&\\
&&\langle P^\prime, S^\prime \vert A_\mu (0) \vert P,S \rangle =
\bar u_{S^\prime} (P^\prime)
\underbrace{\left[
\gamma_\mu \gamma_5\, G^q_A(Q^2) 
+ \gamma_5\,q_\mu\, G^q_P(Q^2) 
\right]}_{\Gamma_\mu^A(P,P^\prime)}
\,U_S(P). \nonumber \\ &&
\end{eqnarray}
Exclusive processes in principle then offer possibilities to measure vector
or axial vector form factors for various flavor currents. 
One can for instance use the $(\gamma Z)$ interference term in electron-nucleon
scattering or $\nu_e p \rightarrow e^+n$ processes to separate the currents
for different flavors.
\begin{quotation}
\small
Note that the normalizations of the densities for a given flavor
imply for the form factor at $Q^2 = 0$,
\[
F^q_1(0) = G_E^q(0) = n_q ,
\]
and one has e.g. $F_1(0) = \sum_q e_q\,n_q = Q$.
Note that $n_q$ is the number of quarks minus antiquarks.
Other form factors at zero momentum transfer just define
some numbers, e.g. 
\[
F^q_2(0) = \kappa_q, \qquad G^q_M(0) = \mu_q, \qquad
G_A^q(0) = g_A^q.
\]
One has e.g. $G_M(0) = \sum_q e_q\,\mu_q$. 
For the axial currents one finds from
$\beta-decay$ a proton-neutron transition element of the axial
current, that using isospin symmetry can be simply converted into
\be
G_A^{p\rightarrow n}(0) = g_A^u - g_A^d = 1.26.
\label{axform}
\ee
\end{quotation}

\section{Quark correlation functions in leptoproduction}

\begin{figure}[t]
\begin{center}
\epsfig{file=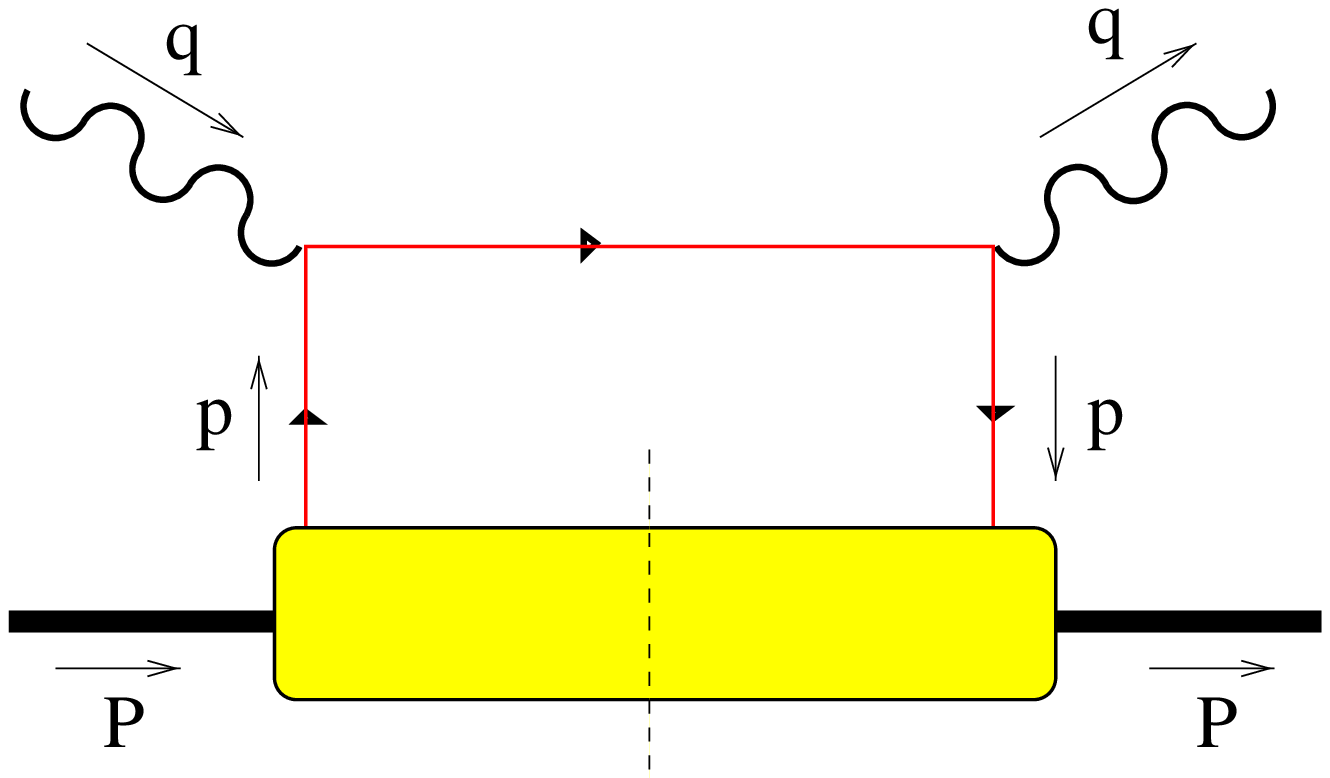,width=4.5cm}
\hspace{2 cm}
\epsfig{file=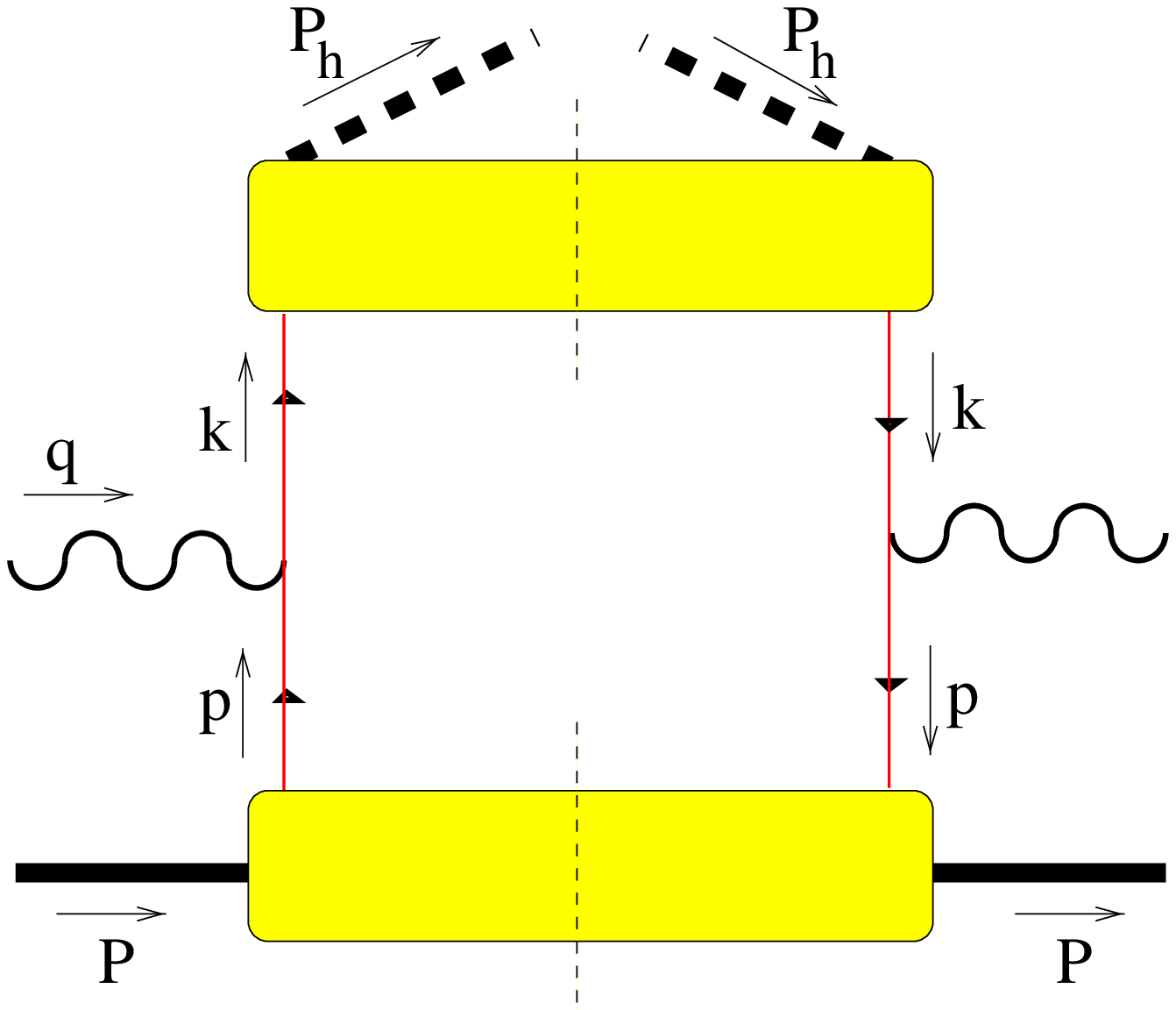,width=4.5cm}
\end{center}
\caption{\label{fig2}
The simplest (parton-level) diagrams representing the squared amplitude
in lepton hadron inclusive scattering (left) and semi-inclusive scattering
(right). In both cases also the diagram with opposite fermion flow has to
be added.}
\end{figure}

Within the framework of QCD and knowing that the photon or $Z^0$ current
couples to the quarks, it is possible to write down a diagrammatic expansion 
for leptoproduction, with in the deep inelastic limit
($Q^2 \rightarrow \infty$) as relevant diagrams only the
ones given in Fig.~\ref{fig2} for
inclusive and 1-particle inclusive scattering respectively.
The expression for ${\cal W}_{\mu\nu}$ can be rewritten as a nonlocal
product of currents and
it is a straightforward exercise to show 
by inserting the currents 
$j_\mu(x) = :\overline \psi(x) \gamma_\mu \psi(x):$ 
that for 1-particle inclusive scattering 
one obtains in tree approximation
\bea
&&
2M{\cal W}_{\mu\nu}( q; P S; P_h S_h ) \nonumber \\
& & \qquad \mbox{} = 
\frac{1}{(2\pi)^4} \int d^4x\ e^{iq\cdot x}\,
\langle P S |:\overline \psi_j (x) (\gamma_\mu)_{jk} \psi_k(x):
\sum_X |X; P_h S_h \rangle \nonumber \\
&&\qquad \qquad \qquad \qquad \qquad \times \langle X; P_h S_h |
:\overline \psi_l(0) (\gamma_\nu)_{li} \psi_i(0):|P S \rangle \nonumber \\
& & \qquad \mbox{} = \frac{1}{(2\pi)^4} \int d^4x\ e^{iq\cdot x}\,
\langle P S |\overline \psi_j (x) \psi_i(0) \vert P S \rangle 
(\gamma_\mu)_{jk}
\nonumber \\
&& \qquad \qquad \qquad  \langle 0 \vert \psi_k(x)  
\sum_X |X; P_h S_h \rangle
\langle X; P_h S_h | \overline \psi_l(0) \vert 0 \rangle (\gamma_\nu)_{li}
\nonumber \\ && \qquad \quad \mbox{}
+ \frac{1}{(2\pi)^4} \int d^4x\ e^{iq\cdot x}\,
\langle P S |\psi_k (x) \overline \psi_l(0) \vert P S \rangle
(\gamma_\nu)_{li} \nonumber \\
&&\qquad \qquad \qquad  \langle 0 \vert \overline \psi_j(x)  \sum_X |X; P_h
S_h \rangle \langle X; P_h S_h |
\psi_i(0) \vert 0 \rangle (\gamma_\mu)_{jk}, \nonumber \\
& & \qquad \mbox{} = 
\int d^4p \,d^4k\,\delta^4(p+q-k) \ \mbox{Tr}\left( \Phi(p) \gamma_\mu
\Delta(k) \gamma_\nu \right)
+ \left\{ \begin{array}{c} q \leftrightarrow -q \\ \mu \leftrightarrow \nu
\end{array} \right\},
\nonumber \\ &&
\label{basic}
\eea
where 
\bea
&& \Phi_{ij}(p) =  \frac{1}{(2\pi)^4} \int d^4\xi\ e^{ip\cdot \xi}\,
\langle P S |\overline \psi_j (0) \psi_i(\xi)|P S \rangle, \nonumber \\
&& \Delta_{kl}(k) = \frac{1}{(2\pi)^4} \int d^4\xi\ e^{ik\cdot \xi}\,
\langle 0 \vert \psi_k(\xi) \sum_X |X; P_h S_h \rangle \langle X; P_h S_h |
\overline \psi_l(0) \vert 0 \rangle. \nonumber  
\eea
Note that in $\Phi$ (quark
production) a summation over colors is assumed, while in $\Delta$ 
(quark decay)
an averaging over colors is assumed. The quantities $\Phi$ and $\Delta$
correspond to the blobs in Fig.~\ref{fig2} and parametrize the soft physics.
Soft refers to all invariants of momenta being small as compared to the hard
scale, i.e. for $\Phi(p)$ one has $p^2 \sim p\cdot P \sim P^2 = M^2 \ll Q^2$.

In general many more diagrams have to be considered in evaluating the
hadron tensors, but in the deep inelastic limit they can be neglected
or considered as corrections to the soft blobs. We return to this later.

As mentioned above,
the relevant structural information for the
hadrons is contained in soft parts (the blobs in Fig.~\ref{fig2})
which represent specific matrix elements of quark fields.
The form of $\Phi$ is constrained by hermiticity, parity and time-reversal
invariance. The quantity depends besides the quark momentum $p$ on the
target momentum $P$ and the spin vector $S$ and one must have
\begin{eqnarray}
&& \mbox{[Hermiticity]} \quad \Rightarrow \quad
\Phi^\dagger (p,P,S) = \gamma_0 \,\Phi(p,P,S)\,\gamma_0 ,
\\
&& \mbox{[Parity]} \quad \Rightarrow \quad
\Phi(p,P,S) = \gamma_0 \,\Phi(\bar p,\bar P,-\bar S)\,\gamma_0 ,
\\
&&\mbox{[Time\ reversal]} \ \Rightarrow 
\ \Phi^\ast(p,P,S) = (-i\gamma_5 C)\,\Phi(\bar p,\bar P,
\bar S)\,(-i\gamma_5 C) ,
\end{eqnarray}
where $C$ = $i\gamma^2 \gamma_0$, $-i\gamma_5 C$= $i\gamma^1\gamma^3$
and $\bar p$ = $(p^0,-\bm p)$.
The most general way to parametrize $\Phi$ using only the constraints
from hermiticity and parity invariance, is~\cite{RS79,TM95}
\bea
\Phi(p,P,S) & = &
M\,A_1 + A_2\,\slash P + A_3 \slash p
+ i\,A_4\,\frac{[\slash P,\slash p]}{2M}
\nonumber \\ & &
+ i\,A_{5}\,(p\cdot S) \gamma_5
+ M\,A_6 \,\slash S \gamma_5
+ A_7\,\frac{(p\cdot S)}{M}\,\slash P \gamma_5
\nonumber \\ & &
+ A_8\,\frac{(p\cdot S)}{M}\,\slash p \gamma_5
+ A_9\,\frac{[\slash P,\slash S]}{2}\,\gamma_5
+ A_{10}\,\frac{[\slash p,\slash S]}{2}\,\gamma_5
\nonumber \\ & &
+ A_{11}\,\frac{(p\cdot S)}{M}\,\frac{[\slash P,\slash p]}{2M}\,\gamma_5
+ A_{12}\,\frac{\epsilon_{\mu \nu \rho \sigma}\gamma^\mu P^\nu
p^\rho S^\sigma}{M},
\label{lorentz}
\eea
where the first four terms do not involve
the hadron polarization vector. Hermiticity requires all the amplitudes
$A_i$ = $A_i(p\cdot P, p^2)$ to be real. The amplitudes $A_4$, $A_5$
and $A_{12}$ vanish when also time reversal invariance applies.

\section{Inclusive scattering}

\subsection{The relevant soft parts}

In order to find out which information in the soft parts 
is important in a hard process one needs to realize
that the hard scale $Q$ leads in a natural way to the use of lightlike
vectors $n_+$ and $n_-$ satisfying $n_+^2 = n_-^2 = 0$ and $n_+\cdot n_-$
= 1. For inclusive scattering one parametrizes the momenta
\[
\left.
\begin{array}{l} q^2 = -Q^2 \\
P^2 = M^2\\
2\,P\cdot q = \frac{Q^2}{\xbj} \\
\end{array} \right\}
\longleftrightarrow \left\{
\begin{array}{l}
q =\ \frac{Q}{\sqrt{2}}\,n_- \ - \ \frac{Q}{\sqrt{2}}\,n_+ 
\\ \mbox{} \\
P = \frac{\xbj M^2}{Q\sqrt{2}}\,n_-
+ \frac{Q}{\xbj \sqrt{2}}\,n_+
\end{array}
\right.
\]
The above are the external momenta. Next turn to the internal
momenta, looking at the left diagram in Fig.~\ref{fig2}. 
In the soft part actually {\em all} momenta, that is $p$ {\em and} $P$ 
have a minus
component that can be neglected compared to that in the hard part,
since otherwise $p\cdot P$ would be hard. Thus because $p$ must have
only a hard plus component, $q$ has two hard components and $k$ being
the current jet also must be soft, i.e. only can have one large
lightcone component, one must have
\begin{eqnarray*}
p &=& \quad\ldots\quad + \frac{Q}{\sqrt{2}}\,n_+ ,
\\
q &=& \frac{Q}{\sqrt{2}}\,n_- - \frac{Q}{\sqrt{2}}\,n_+ ,
\\
p+q = k &=& \frac{Q}{\sqrt{2}}\,n_- + \quad\ldots\ .
\end{eqnarray*}
where the \ldots parts indicate (negligible) $1/Q$ terms.

Also the transverse component is not relevant for the hard part.
One thus sees that for inclusive scattering the only relevant
dependence of the soft part is the $p^+$ dependence. Moreover,
the above requirements on the internal momenta already indicate 
that the lightcone fraction $x = p^+/P^+$ must be equal to $\xbj$.
This will come out when we do the actual calculation in one
of the next sections.

The minus component $p^- \equiv p\cdot n_+$ and transverse components 
thus can be integrated over restricting the nonlocality in $\Phi(p)$.
The relevant soft part then is some Dirac trace
of the quantity~\cite{Soper77,Jaffe83}
\bea
\Phi_{ij}(x) & = & 
\int dp^-\,d^2p_\st \ \Phi_{ij}(p,P,S)
\nonumber \\
& = &
\left. \int \frac{d\xi^-}{2\pi}\ e^{ip\cdot \xi}
\,\langle P,S\vert \overline \psi_j(0) \psi_i(\xi)
\vert P,S\rangle \right|_{\xi^+ = \xi_\st = 0},
\eea
depending on the lightcone fraction $x = p^+/P^+$. 
To be precise one puts in the full form for the quark momentum,
\be
p = x\,P^+n_+ + \frac{p^2 + \bm p_\st^2}{2x\,P^+}\,n_- + p_\st,
\label{quarkmom}
\ee
and performs the integration over $\Phi(p)$ using
\be
\int dp^-\,d^2p_\st \ldots = \frac{\pi}{P^+}\int d(p\cdot P)\,dp^2 
\ldots \ .
\ee
When one wants to calculate the leading order in $1/Q$ for a hard 
process, one only needs to look at leading parts in $M/P^+$ because
$P^+ \propto Q$ (see opening paragraph of this section)~\cite{JJ92}. 
In this case that turns out
to be the part proportional to $(M/P^+)^0$, 
\be
\Phi(x) =
\frac{1}{2}\,\Biggl\{
f_1(x)\,\nslash_+
+ \lambda\,g_1(x)\, \gamma_5\,\nslash_+
+ h_1(x)\,\frac{\gamma_5\,[\Sslash_\perp,\nslash_+]}{2}\Biggr\}
+ {\cal O}\left(\frac{M}{P^+}\right)
\ee
The precise expression of the functions $f_1(x)$, etc. as integrals
over the amplitudes can be easily written down.

\subsection{Calculating the inclusive cross section}

Using field theoretical methods the left diagram
in Fig.~\ref{fig2} can now be calculated. 
Omitting the sum over flavors ($\sum_a$), 
the quark charges $e_a^2$ and the
$(q\leftrightarrow -q, \mu\leftrightarrow \nu)$ 'antiquark' diagram,
the symmetric part of the hadron tensor the result is
\be
2M\,W^{\mu \nu}(P,q) 
= \int dp^-\,dp^+\,d^2p_\perp \ \mbox{Tr}\left(\Phi(p)
\,\gamma^\mu \Delta(p+q) \gamma^\nu\right),
\ee
where 
\be
\Delta(k) = (\slash k + m)\,\delta(k^2-m^2) 
\approx \frac{\slash n_-}{2}\,\delta(k^+) ,
\ee
and in the approximation anything proportional to $1/Q^2$ has
been neglected. One obtains
\begin{eqnarray}
2M\,W_S^{\mu \nu}(P,q) 
&=& \int dp^-\,dp^+\,d^2p_\perp \ \frac{1}{2}\,\mbox{Tr}\left(\Phi(p)
\,\gamma^\mu \,\gamma^+\,\gamma^\nu\right)
\, \delta ( p^+ + q^+) \nonumber \\
&= & -g_\perp^{\mu\nu}\ \left. \mbox{Tr}\left(\gamma^+\,\Phi(x) \right)
\right|_{x = \xbj} 
\nonumber \\
& = & -g_\perp^{\mu\nu}\,f_1(\xbj).
\label{inclcalc}
\end{eqnarray}
Antiquarks arise from the diagram with opposite fermion flow, 
proportional to
$\mbox{Tr}\left(\overline\Phi(p) 
\,\gamma^\nu \overline\Delta(k) \gamma^\mu\right)$
with
\be
\overline \Phi_{ij}(p) =
\frac{1}{(2\pi)^4} \int d^4\xi\ e^{-ip\cdot \xi}
\,\langle P S |\psi_i (\xi) \overline \psi_j(0)|P S \rangle .
\ee
The {\em proper} definition of antiquark distributions starts from
$\Phi^c(x)$ containing antiquark distributions $\bar f_1(x)$, etc.
The quantity $\Phi^c(p)$ is obtained from $\Phi(p)$ after
the replacement of $\psi$ by $\psi^C = C\overline\psi^T$.
One then finds $\overline \Phi(p)$ = $-C(\Phi^c)^TC^\dagger$, i.e.
one has to be aware of sign differences. 
Symmetry relations between quark and antiquark relations
can be obtained using the anticommutation relations for fermions,
giving $\overline \Phi_{ij}(p) = - \Phi_{ij}(-p)$. One finds that
$\bar f_1(x) = -f_1(-x)$, $\bar g_1(x) = g_1(-x)$, and
$\bar h_1(x) = -h_1(-x)$.
Finally, after including the flavor summation and the quark 
charges squared one can compare the result with
Eq.~\ref{param} to obtain for the structure function
\be
2F_1(\xbj) = \sum_a e_a^2\,\left(f_1^a(\xbj) + f_1^{\bar a}(\xbj)\right),
\ee
while $F_L(\xbj) = 0$ (Callan-Gross relation).

The antisymmetric part of $W^{\mu\nu}$ in the above calculation is
left as an exercise. The answer is
\begin{eqnarray}
2M\,W_A^{\mu \nu}(P,q) 
&=& i\,\epsilon_\perp^{\mu\nu}\,g_1(\xbj),
\end{eqnarray}
which after inclusion of antiquarks, flavor summation gives after
comparison with Eq.~\ref{wanti} 
\be
2g_1(\xbj) = \sum_a e_a^2\,\left(g_1^a(\xbj) + g_1^{\bar a}(\xbj)\right).
\ee

\subsection{Interpretation of the functions}

The functions $f_1$, $g_1$ and $h_1$ can be obtained from the 
correlator $\Phi(x)$ after tracing with the appropriate 
Dirac matrix,
\bea
f_1(x) & = &
\left. \int \frac{d\xi^-}{4\pi}\ e^{ip\cdot \xi}
\,\langle P,S\vert \overline \psi(0) \gamma^+ \psi(\xi)
\vert P,S\rangle \right|_{\xi^+ = \xi_\st = 0},
\\
\lambda\,g_1(x) & = &
\left. \int \frac{d\xi^-}{4\pi}\ e^{ip\cdot \xi}
\,\langle P,S\vert \overline \psi(0) \gamma^+\gamma_5 \psi(\xi)
\vert P,S\rangle \right|_{\xi^+ = \xi_\st = 0},
\\
S_\st^i\,h_1(x) & = &
\left. \int \frac{d\xi^-}{4\pi}\ e^{ip\cdot \xi}
\,\langle P,S\vert \overline \psi(0) \,i\sigma^{i+}\gamma_5\,\psi(\xi)
\vert P,S\rangle \right|_{\xi^+ = \xi_\st = 0},
\eea
By introducing {\em good} and {\em bad} fields 
$\psi_\pm \equiv \frac{1}{2}\gamma^\mp\gamma^\pm \psi$, one
sees that $f_1$ can be rewritten as
\bea
f_1(x) & = &
\left. \int \frac{d\xi^-}{2\pi\sqrt{2}}\ e^{ip\cdot \xi}
\,\langle P,S\vert \psi^\dagger_+(0) \psi_+(\xi)
\vert P,S\rangle \right|_{\xi^+ = \xi_\st = 0}
\nonumber \\
& = & \frac{1}{\sqrt{2}}\sum_n \left| \langle P_n\vert 
\psi_+\vert P\rangle\right|^2\,\delta\left(P_n^+ - (1-x)P^+\right) ,
\eea
i.e. it is a quark lightcone momentum distribution. For the functions
$g_1$ and $h_1$ one needs in addition the projectors on quark chirality
states, $P_{R/L} = \frac{1}{2}(1\pm \gamma_5)$, and on quark transverse
spin states~\cite{Artru,JJ92}, 
$P_{\uparrow/\downarrow} = \frac{1}{2}(1\pm \gamma^i
\gamma_5)$ to see that
\bea
&& 
f_1(x) = f_{1R}(x) + f_{1L}(x) = f_{1\uparrow}(x) + f_{1\downarrow}(x),
\\ &&
g_1(x) = f_{1R}(x) - f_{1L}(x), 
\\ &&
h_1(x) = f_{1\uparrow}(x) - f_{1\downarrow}(x).
\eea
One sees some trivial bounds such as $f_1(x) \ge 0$ and
$\vert g_1(x)\vert \le f_1(x)$.
Since $P_n^+ \le 0$ and sees $x \le 1$. From the antiquark
distribution $\bar f_1(x)$ and its relation to $f_1(x)$ one
obtains $x \ge -1$, thus the support of the functions is
$-1 \le x \le 1$.

\subsection{Bounds on the distribution functions}

The trivial bounds on the distribution functions
($\vert h_1(x)\vert \le f_1(x)$ and $\vert g_1(x)\vert 
\le f_1(x)$) can be sharpened. For instance
one can look explicitly at the structure in Dirac space of the 
correlation function $\Phi_{ij}$. Actually, we will look at the
correlation functions $(\Phi\,\gamma_0)_{ij}$, which involves 
at leading order matrix elements $\psi_{+j}^\dagger (0)\psi_{+i}(\xi)$.
One has in Weyl representation
($\gamma^0 = \rho^1$,
$\gamma^i = -i\rho^2\sigma^i$,
$\gamma_5 = i\gamma^0\gamma^1\gamma^2\gamma^3 = \rho^3$)
the matrices
\begin{eqnarray*}
P_+ & = &
\left\lgroup \begin{array}{rrrr}
1 & 0 & 0 & 0 \\
0 & 0 & 0 & 0 \\
0 & 0 & 0 & 0 \\
0 & 0 & 0 & 1
\end{array}\right\rgroup, \quad
\\
P_+\gamma_5 & = &
\left\lgroup \begin{array}{rrrr}
1 & 0 & 0 & 0 \\
0 & 0 & 0 & 0 \\
0 & 0 & 0 & 0 \\
0 & 0 & 0 & -1
\end{array}\right\rgroup, \quad
P_+\gamma^1\gamma_5 =
\left\lgroup \begin{array}{rrrr}
0 & 0 & 0 & 1 \\
0 & 0 & 0 & 0 \\
0 & 0 & 0 & 0 \\
1 & 0 & 0 & 0
\end{array}\right\rgroup .
\end{eqnarray*}
The good projector only leaves two (independent) Dirac spinors, one
righthanded (R), one lefthanded (L). 
On this basis of good R and L spinors the for hard scattering processes 
relevant matrix $(\Phi\slash n_-)$ is given by
\be
(\Phi\,\slash n_-)_{ij}(x) =
\left\lgroup \begin{array}{cc}
f_1 + \lambda\,g_1 &  (S_\st^1-i\,S_\st^2)\,h_1 \\
& \\
(S_\st^1+i\,S_\st^2)\,h_1 & f_1 - \lambda\,g_1
\end{array}\right\rgroup
\ee
One can also turn the $S$-dependent correlation function $\Phi$
defined in analogy with $W(q,P,S)$ in Eq.~\ref{defS}
into a matrix in the nucleon spin space. If
\bea
\Phi (x; P,S) &=& \Phi_O + \lambda\,\Phi_L + S_\st^1\,\Phi_\st^1
+ S_\st^2\,\Phi_\st^2 ,
\eea
then one has on the basis of spin 1/2 target states with $\lambda = +1$
and $\lambda = -1$ respectively
\be
\Phi_{ss^\prime}(x) =
\left\lgroup \begin{array}{cc}
\Phi_O + \Phi_L & \Phi_\st^1 - i\,\Phi_\st^2 \\
& \\
\Phi_\st^1 + i\,\Phi_\st^2 & \Phi_O - \Phi_L \\
\end{array}\right\rgroup
\ee
The matrix relevant for bounds is the matrix $M$ = $(\Phi \slash n_-)^T$
(for this matrix one has $v^\dagger M v \ge 0$ for any direction $v$).
On the basis $+R$, $-R$, $+L$ and $-L$ it becomes
\be
(\Phi(x)\,\slash n_-)^T =
\left\lgroup \begin{array}{cccc}
f_1 + g_1 & 0 & 0 & 2\,h_1 \\
& \\
0 & f_1 - g_1 & 0 & 0 \\
& \\
0 & 0 & f_1 - g_1 & 0 \\
& \\
2\,h_1 & 0 & 0 & f_1 + g_1
\end{array}\right\rgroup .
\ee
Of this matrix any diagonal matrix element must always be positive, 
hence the eigenvalues must be positive, which  gives a bound on the
distribution functions stronger than the trivial bounds, namely
\be
\vert h_1(x)\vert \le \frac{1}{2}\left( f_1(x) + g_1(x)\right)
\ee
known as the Soffer bound~\cite{Soffer}.

\subsection{Sum rules}

For the functions appearing in the soft parts, and thus also for the
structure functions, one can derive sum rules. Starting with
the traces defining the quark distributions,
\begin{eqnarray*}
f_1(x) & = &
\left. \int \frac{d\xi^-}{4\pi}\ e^{ip\cdot \xi}
\,\langle P,S\vert \overline \psi(0) \gamma^+ \psi(\xi)
\vert P,S\rangle \right|_{\xi^+ = \xi_\st = 0},
\\
g_1(x) & = &
\left. \int \frac{d\xi^-}{4\pi}\ e^{ip\cdot \xi}
\,\langle P,S\vert \overline \psi(0) \gamma^+\gamma_5 \psi(\xi)
\vert P,S\rangle \right|_{\xi^+ = \xi_\st = 0},
\end{eqnarray*}
and integrating over $x = p^+/P^+$ one obtains (using symmetry relation
as indicated above to eliminate antiquarks $\bar f_1$),
\bea
\int_0^1 dx\,\left( f_1(x) - \bar f_1(x)\right)
= \int_{-1}^1 dx\ f_1(x) 
= \frac{\langle P,S\vert \overline \psi(0) \gamma^+ \psi(0)
\vert P,S\rangle }{2P^+},
\eea
which as we have seen in the section on elastic scattering is nothing
else than a form factor at zero momentum transfer, i.e. the number
of quarks of that particular flavor. Similarly one finds the sum rule
\bea
\int_0^1 dx\,\left( g_1(x) + \bar g_1(x)\right)
= \int_{-1}^1 dx\ g_1(x) 
= \frac{\langle P,S\vert \overline \psi(0) \gamma^+\gamma_5 \psi(0)
\vert P,S\rangle }{2P^+},
\label{axsumr}
\eea
which precisely is the axial charge $g_A$ for a particular quark flavor.
These sum rules for the quark distributions underly the sum rules for
the structure functions, e.g. the Bjorken sum rule following from
Eq.~\ref{axsumr} and Eq.~\ref{axform}.
\be
\int_0^1 d\xbj\ \left(g_1^p(\xbj,Q^2)-g_1^n(\xbj,Q^2)\right)
= \frac{1}{6}\left(g_A^u - g_A^d\right) 
= \frac{1}{6}\,G_A^{p\rightarrow n}(0).
\ee

\section{1-particle inclusive scattering}

\subsection{The relevant distribution functions}

For 1-particle inclusive scattering one parametrizes the momenta
\[
\left.
\begin{array}{l} q^2 = -Q^2 \\
P^2 = M^2\\
P_h^2 = M_h^2 \\
2\,P\cdot q = \frac{Q^2}{\xbj} \\
2\,P_h\cdot q = -z_h\,Q^2
\end{array} \right\}
\longleftrightarrow \left\{
\begin{array}{l}
P_h = \frac{z_h\,Q}{\sqrt{2}}\,n_-
+ \frac{M_h^2}{z_h\,Q\sqrt{2}}\,n_+
\\ \mbox{} \\
q =\ \frac{Q}{\sqrt{2}}\,n_- \ - \ \frac{Q}{\sqrt{2}}\,n_+\ +\ q_T
\\ \mbox{} \\
P = \frac{\xbj M^2}{Q\sqrt{2}}\,n_-
+ \frac{Q}{\xbj \sqrt{2}}\,n_+
\end{array}
\right.
\]
Note that this works for socalled current fragmentation,
in which case the produced hadron is {\em hard} with respect
to the target momentum, i.e. $P\cdot P_h \sim Q^2$.
The minus component $p^-$ is irrelevant in the lower
soft part, while the plus component $k^+$ is irrelevant in the upper
soft part. Note that after the choice of $P$ and $P_h$ one can no
longer omit a transverse component in the other vector, in this
case the momentum transfer $q$. This is
precisely the vector $q_T$ introduced earlier in the discussion
of the structure functions for 1-particle inclusive leptoproduction.
One immediately sees that one can no longer simply integrate
over the transverse component of the quark momentum, defined in
Eq.~\ref{quarkmom}.

At this point it turns out that the most convenient way to describe
the spin vector of the target is via an expansion of the form
\bea
S^\mu =
-\lambda\,\frac{M\xbj}{Q\sqrt{2}}\,n_- 
+ \lambda\,\frac{Q}{M\xbj\sqrt{2}}\,n_+ + S_\st.
\eea
One has up to ${\cal O}(1/Q^2)$ corrections
$\lambda \approx M\,(S\cdot q)/(P\cdot q)$ and 
$S_\st \approx S_\perp$. For a pure state one has 
$\lambda^2 + \bm S_\st^2 = 1$, 
in general this quantity being less or equal than one.

\begin{figure}[t]
\leavevmode
\begin{center}
\begin{minipage}{9.0cm}
\epsfig{file=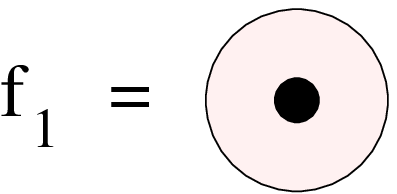,width=1.8cm}
\hspace{2.5 cm}
\epsfig{file=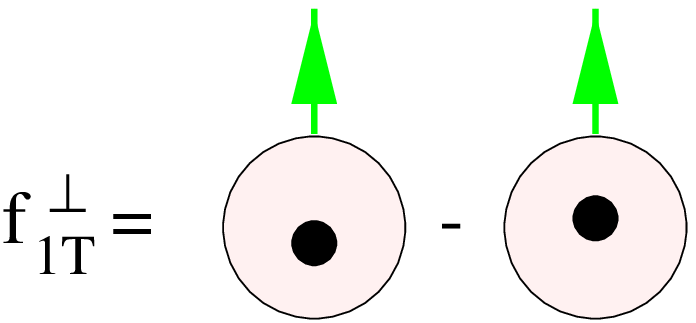,width=3.1cm}
\\[0.2cm]
\epsfig{file=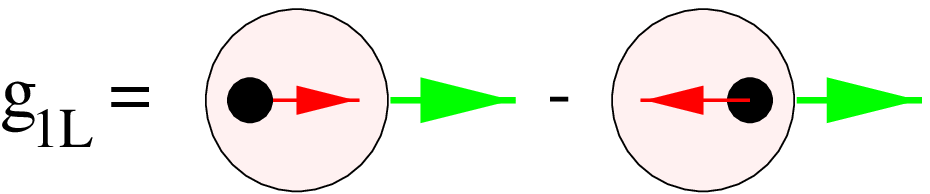,width=4.2cm}
\hspace{0.5 cm}
\epsfig{file=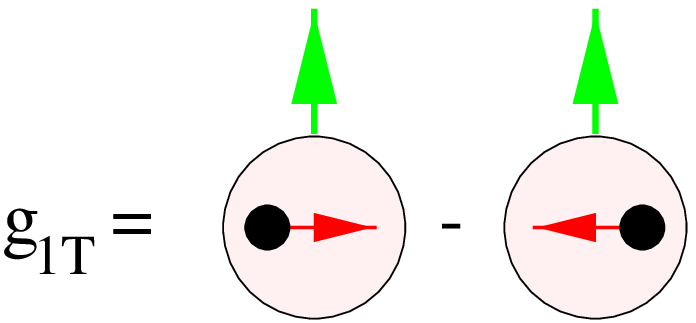,width=3.1cm}
\\[0.2cm]
\epsfig{file=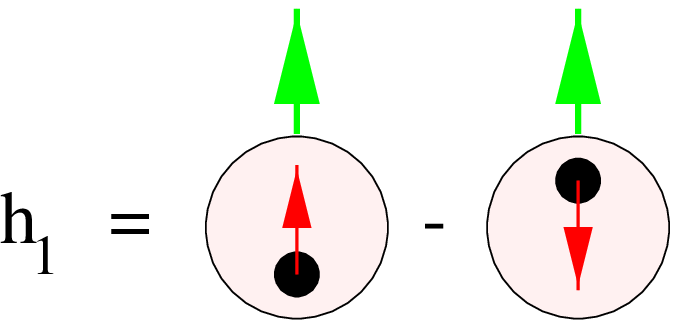,width=3.1cm}
\hspace{1.3cm}
\epsfig{file=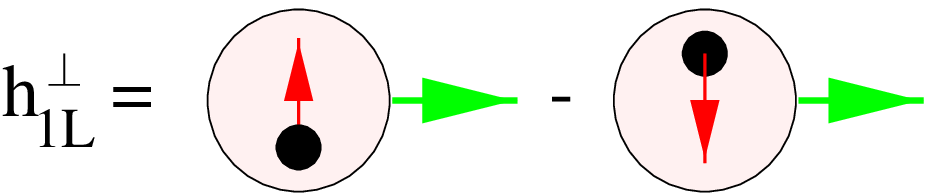,width=4.2cm}
\\[0.6cm]
\epsfig{file=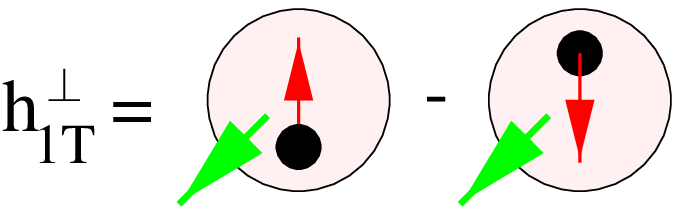,width=3.1cm}
\hspace{1.3 cm}
\epsfig{file=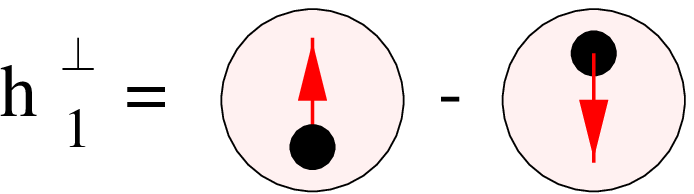,width=3.1cm}
\end{minipage}
\end{center}
\caption{
Interpretation of the functions in the leading
Dirac traces of $\Phi$.
\label{fig3}
}
\end{figure}

The soft part to look at is
\be
\Phi(x,\bm p_T) =
\left. \int \frac{d\xi^-d^2\bm \xi_T}{(2\pi)^3}\ e^{ip\cdot \xi}
\,\langle P,S\vert \overline \psi(0) \psi(\xi)
\vert P,S\rangle \right|_{\xi^+ = 0}.
\ee
For the leading order results, it is parametrized as
\be
\Phi(x,\bm p_\st)  =
\Phi_O(x,\bm p_\st) + \Phi_L(x,\bm p_\st) + \Phi_T(x,\bm p_\st),
\ee
with the parts involving unpolarized
targets (O), longitudinally polarized targets (L) and transversely
polarized targets (T) up to parts proportional to $M/P^+$ given by
\bea
\Phi_O(x,\bm p_\st) & = &
\frac{1}{2} \Biggl\{
f_1(x,\bm p_\st)\,\nslash_+
+ h_1^\perp(x,\bm p_\st)\,\frac{i\,[\pslash_\st,\nslash_+]}{2M}
\Biggr\}
\\ 
\Phi_L(x,\bm p_\st) & = &
\frac{1}{2} \Biggl\{
\lambda\,g_{1L}(x,\bm p_\st)\,\gamma_5\,\nslash_+
+ \lambda\,h_{1L}^\perp(x,\bm p_\st)
\frac{\gamma_5\,[\pslash_\st,\nslash_+]}{2M}
\Biggr\}
\\ 
\Phi_T(x,\bm p_\st)  & = & 
\frac{1}{2} \Biggl\{
f_{1T}^\perp(x,\bm p_\st)\, \frac{\epsilon_{\mu \nu \rho \sigma}
\gamma^\mu n_+^\nu p_\st^\rho S_\st^\sigma}{M}
\nonumber \\ & &\mbox{}
+ \frac{\bm p_\st\cdot\bm S_\st}{M}\,g_{1T}(x,\bm p_\st)
\,\gamma_5\,\nslash_+
+ h_{1T}(x,\bm p_\st)\,\frac{\gamma_5\,[\Sslash_\st,\nslash_+]}{2}
\nonumber \\ & &\mbox{}
+ \frac{\bm p_\st\cdot\bm S_\st}{M}\,h_{1T}^\perp(x,\bm p_\st)\,
\frac{\gamma_5\,[\pslash_\st,\nslash_+]}{2M}
\Biggr\}.
\eea
All functions appearing here have a natural interpretation
as densities. This is seen as discussed before for the 
$\bm p_\st$-integrated functions. Now it includes densities such as 
the density of
longitudinally polarized quarks in a transversely polarized nucleon
($g_{1T}$) and the density of transversely polarized quarks in
a longitudinally polarized nucleon ($h_{1L}^\perp$). The interpretation
of all functions is illustrated in Fig.~\ref{fig3}. 

Several functions vanish from the soft part upon integration
over $p_\st$. Actually we will find that particularly interesting
functions survive when one integrates over $\bm p_\st$ weighting
with $p_\st^\alpha$, e.g.
\bea
\Phi_\partial^\alpha (x) &\equiv&
\int d^2 p_\st\,\frac{p_\st^\alpha}{M} \,\Phi(x,\bm p_\st)
\nonumber \\ & = & 
\frac{1}{2}\,\Biggl\{
-g_{1T}^{(1)}(x)\,S_\st^\alpha\,\slash n_+\gamma_5 
-\lambda\,h_{1L}^{\perp (1)}(x)
\,\frac{[\gamma^\alpha,\slash n_+]\gamma_5}{2}
\nonumber \\
&&\quad \mbox{} 
-{f_{1T}^{\perp (1)}}
\,\epsilon^{\alpha}_{\ \ \mu\nu\rho}\gamma^\mu n_-^\nu {S_\st^\rho}
- {h_1^{\perp (1)}}
\,\frac{i[\gamma^\alpha, \nslash_+]}{2}\Biggr\},
\label{Phid}
\eea
where we define $\bm p_\st^2/2M^2$-moments as
\be
g_{1T}^{(1)}(x) = \int d^2p_\st\ \frac{\bm p_\st^2}{2M^2}
\,g_{1T}(x,\bm p_\st),
\ee
and similarly the other functions.
The functions
$h_1^\perp$ and $f_{1T}^\perp$ are T-odd, vanishing if T-reversal
invariance can be applied to the matrix element. For $p_\st$-dependent
correlation functions, matrix elements involving gluon fields at
infinity (gluonic poles~\cite{bmt}) can for instance prevent 
application of T-reversal invariance. 
The functions describe the possible
appearance of unpolarized quarks in a transversely polarized nucleon
($f_{1T}^\perp$) or transversely polarized quarks in an unpolarized
hadron ($h_1^\perp$) and lead to single-spin asymmetries in various
processes~\cite{Sivers90,Anselmino95}.
The interpretation of these functions is also
illustrated in Fig.~\ref{fig3}.
Of course just integrating $\Phi(x,p_\st)$ over $p_\st$ gives the
result used in inclusive scattering with $f_1(x)$ = $\int d^2p_\st
\ f_1(x,p_\st)$, $g_1(x)$ = $g_{1L}(x)$ and $h_1(x)$ = $h_{1T}(x)
+ h_{1T}^{\perp (1)}(x)$. We note that the function $h_{1T}^{\perp (2)}$
appears after weighting with $p_\st^\alpha p_\st^\beta$.

\subsection{The relevant fragmentation functions}

Just as for the distribution functions one can perform an analysis of
the soft part describing the quark fragmentation.
One needs~\cite{CS82} 
\be
\Delta_{ij}(z,\bt k) =
\left. \sum_X \int \frac{d\xi^+d^2\bm \xi_\st}{(2\pi)^3} \,
e^{ik\cdot \xi} \,Tr  \langle 0 \vert \psi_i (\xi) \vert P_h,X\rangle
\langle P_h,X\vert\overline \psi_j(0) \vert 0 \rangle
\right|_{\xi^- = 0}.
\ee
For the production of unpolarized hadrons $h$ in hard processes one needs 
to leading order in $1/Q$ the correlation function,
\be
\Delta_O(z,\bm k_\st) =
z\,D_1(z,\bm k^\prime_\st)\,\nslash_-
+ z\,H_1^\perp(z,\bm k^\prime_\st)\,\frac{i\,[\kslash_\st,\nslash_-]}{2M_h}
+ {\cal O}\left(\frac{M_h}{P_h^-}\right).
\ee
when we limit ourselves to an unpolarized or spin 0 final state hadron.
The arguments of the fragmentation functions $D_1$ and $H_1^\perp$ are
$z$ = $P_h^-/k^-$ and $\bm k^\prime_\st$ = $-z\bt k$. The first
is the (lightcone) momentum fraction of the produced hadron, the second
is the transverse momentum of the produced hadron with respect to the quark.
The fragmentation function $D_1$ is the equivalent of the distribution
function $f_1$. It can be interpreted as the probability of finding a
hadron $h$ in a quark. 
The function $H_1^\perp$, interpretable as the difference in
production probabilities of unpolarized hadrons from a transversely
polarized quark depending on transverse momentum, is allowed 
because of the non-applicability of time reversal 
invariance~\cite{Collins93}. 
This is natural for the fragmentation functions~\cite{HHK83,JJ93} 
because of the  appearance of out-states
$\vert P_h, X\rangle$ in the definition of $\Delta$, in contrast
to the plane wave states appearing in $\Phi$.
After $\bt k$-averaging one is left with the functions
$D_1(z)$ and the $\bm k_\st^2/2M^2$-weighted result $H_1^{\perp (1)}(z)$.

\subsection{The semi-inclusive cross section}

After the analysis of the soft parts, the next step is to find
out how one obtains the information on the various correlation functions
from experiments, in this particular case in lepton-hadron scattering
via one-photon exchange as discussed before.
To get the leading order result for semi-inclusive scattering it is
sufficient to compute the diagram in Fig.~\ref{fig2} (right)
by using QCD and QED Feynman rules in the hard part and the
matrix elements $\Phi$ and $\Delta$ for the soft parts, parametrized in
terms of distribution and fragmentation functions. The most
well-known results for leptoproduction are:
\bea
&&\frac{d\sigma_{OO}}{d\xbj\,dy\,dz_h}
= \frac{2\pi \alpha^2\,s}{Q^4}\,\sum_{a,\bar a} e_a^2
\left\lgroup 1 + (1-y)^2\right\rgroup \xbj {f^a_1}(\xbj)\,{ D^a_1}(z_h)
\\ && \frac{d\sigma_{LL}}{d\xbj\,dy\,dz_h}
= \frac{2\pi \alpha^2\,s}{Q^4}\,{\lambda_e\,\lambda}
\,\sum_{a,\bar a} e_a^2\  y (2-y)\  \xbj {g^a_1}(\xbj)\,{D^a_1}(z_h)
\eea
The indices attached to the cross section refer to polarization
of lepton (O is unpolarized, L is longitudinally polarized) and
hadron (O is unpolarized, L is longitudinally polarized, T is 
transversely polarized). Note that the result is a weighted sum
over quarks and antiquarks involving the charge $e_a$ squared.
Comparing with well-known formal expansions of the cross section in
terms of structure functions one can simply identify these. 
For instance the above result for unpolarized scattering (OO) 
shows that after averaging over azimuthal angles,
only one structure function survives if we work at order $\alpha_s^0$ 
and at leading order in $1/Q$.

As we have seen, in 1-particle inclusive unpolarized leptoproduction
in principle four structure functions appear, two of them containing azimuthal
dependence of the form $\cos (\phi_h^\ell)$ and $\cos (2\phi_h^\ell)$.
The first one only appears at order $1/Q$~\cite{LM94}, the second one
even at leading order but only in the case of the existence
of nonvanishing T-odd distribution functions. To be specific if we
define weighted cross section such as
\be
\int d^2\bm q_{T}\,\frac{Q_{T}^2}{MM_h} \,\cos(2\phi_h^\ell)
\,\frac{d\sigma_{{OO}}}{d\xbj\,dy\,dz_h\,d^2\bm q_{T}}
\equiv
\left< \frac{Q_{T}^2}{MM_h} \,\cos(2\phi_h^\ell)\right>_{OO}
\ee
we obtain the following asymmetry,.
\be
\left<
\frac{Q_{T}^2}{MM_h} \,\cos(2\phi_h^\ell)\right>_{OO}
= \frac{16\pi \alpha^2\,s}{Q^4}
\,(1-y)\,\sum_{a,\bar a} e_a^2
\,\xbj\,{h_{1}^{\perp(1)a}}(\xbj) H_1^{\perp (1)a}.
\ee
In lepton-hadron scattering this asymmetry requires T-odd distribution
functions and therefore most likely is absent or very small. In
$e^+e^-$ annihilation, however, a $\cos 2\phi$ asymmetry between 
produced particles (e.g. pions) in opposite jets involves two
very likely nonvanishing fragmentation functions $H_1^\perp$ and
$\overline H_1^\perp$. Indications for the presence of these
fragmentation functions have been found in LEP data\cite{Efremov}.

For polarized targets, several azimuthal asymmetries arise already
at leading order. For example the following possibilities were
investigated in Refs~\cite{KM96,Collins93,Kotzinian95,TM95b}.
\bea
&&
\left< \frac{Q_\st}
{M} \,\cos(\phi_h^\ell-\phi_S^\ell)\right>_{LT} =
\nonumber \\ && \qquad
\frac{2\pi \alpha^2\,s}{Q^4}\,{\lambda_e\,\vert \bt S \vert}
\,y(2-y)\sum_{a,\bar a} e_a^2
\,\xbj\,{g_{1T}^{(1)a}}(\xbj) {D^a_1}(z_h),
\label{asbas}
\\
&&
\left< \frac{Q_\st^2}{MM_h}
\,\sin(2\phi_h^\ell)\right>_{OL} =
\nonumber \\ && \qquad
-\frac{4\pi \alpha^2\,s}{Q^4}\,{\lambda}
\,(1-y)\sum_{a,\bar a} e_a^2
\,\xbj\,{h_{1L}^{\perp(1)a}}(\xbj) {H_1^{\perp(1)a}}(z_h),
\label{as2}
\\
&&
\left< \frac{Q_\st}{M_h}
\,\sin(\phi_h^\ell+\phi_S^\ell)\right>_{OT} =
\nonumber \\ && \qquad
\frac{4\pi \alpha^2\,s}{Q^4}\,{\vert \bt S \vert}
\,(1-y)\sum_{a,\bar a} e_a^2
\,\xbj\,{h_1^a}(\xbj) {H_1^{\perp(1)a}}(z_h).
\label{finalstate}
\eea
The latter two are single spin asymmetries involving the fragmentation
function $H_1^{\perp (1)}$. The last one was the asymmetry proposed by
Collins~\cite{Collins93} as a way to access the transverse spin 
distribution function $h_1$ in pion production. 
Note, however, that in using the
azimuthal dependence one needs to be very careful. For instance, besides
the $<\sin (\phi_h^\ell + \phi_S^\ell)>_{OT}$, one also finds at leading 
order a $<\sin (3\phi_h^\ell - \phi_S^\ell)>_{OT}$ asymmetry which is 
proportional to $h_{1T}^{\perp (2)}\,H_1^{\perp (1)}$~\cite{TM95b}.

\section{Inclusion of subleading contributions}

\subsection{Subleading inclusive leptoproduction}

If one proceeds up to order $1/Q$ one also needs terms in
the parametrization of the soft part proportional to
$M/P^+$. Limiting ourselves to the $\bm p_\st$-integrated 
correlations one needs
\begin{eqnarray}
\Phi(x) & = & 
\frac{1}{2}\,\Biggl\{
f_1(x)\,\nslash_+
+ \lambda\,g_1(x)\, \gamma_5\,\nslash_+
+ h_1(x)\,\frac{\gamma_5\,[\Sslash_\st,\nslash_+]}{2}\Biggr\}
\nonumber \\ 
& + & \frac{M}{2P^+}\Biggl\{
e(x) + g_T(x)\,\gamma_5\,\Sslash_\st
+ \lambda\,h_L(x)\,\frac{\gamma_5\,[\nslash_+,\nslash_-]}{2} \Biggr\}
\nonumber \\
& + & \frac{M}{2P^+}\Biggl\{
-\lambda\,e_L(x)\,i\gamma_5
- f_T(x)\,\epsilon_\st^{\rho\sigma}\gamma_\rho S_{\st\sigma}
+ h(x)\,\frac{i\,[\nslash_+,\nslash_-]}{2} \Biggr\}.
\end{eqnarray}
The last set of three terms proportional to $M/P^+$ vanish when
time-reversal invariance applies.

Actually in the calculation of the cross section one has to be
careful. Let us use inclusive scattering off a transversely polarized
nucleon (transverse means $\vert \bm S_\perp\vert = 1$ in 
Eq.~\ref{inclspin}) as an example. The hadronic tensor is zero in
leading order in $1/Q$. At order $1/Q$ one obtains from the handbag
diagram a contribution
\be
2M\,W_{A (a)}^{\mu\nu}({q,P,S_\st}) =
i\,\frac{2M}{Q}
\,\hat t_{\mbox{}}^{\,[\mu}\epsilon_\perp^{\nu ]\rho} {S_{\perp\rho}}
\,\left( g_{1T}^{(1)}(\xbj) - \frac{m}{M}\,h_1(\xbj)\right).
\ee
It shows that one must be very careful with the integration over $p_\st$.

\begin{figure}[t]
\begin{center}
\epsfig{file=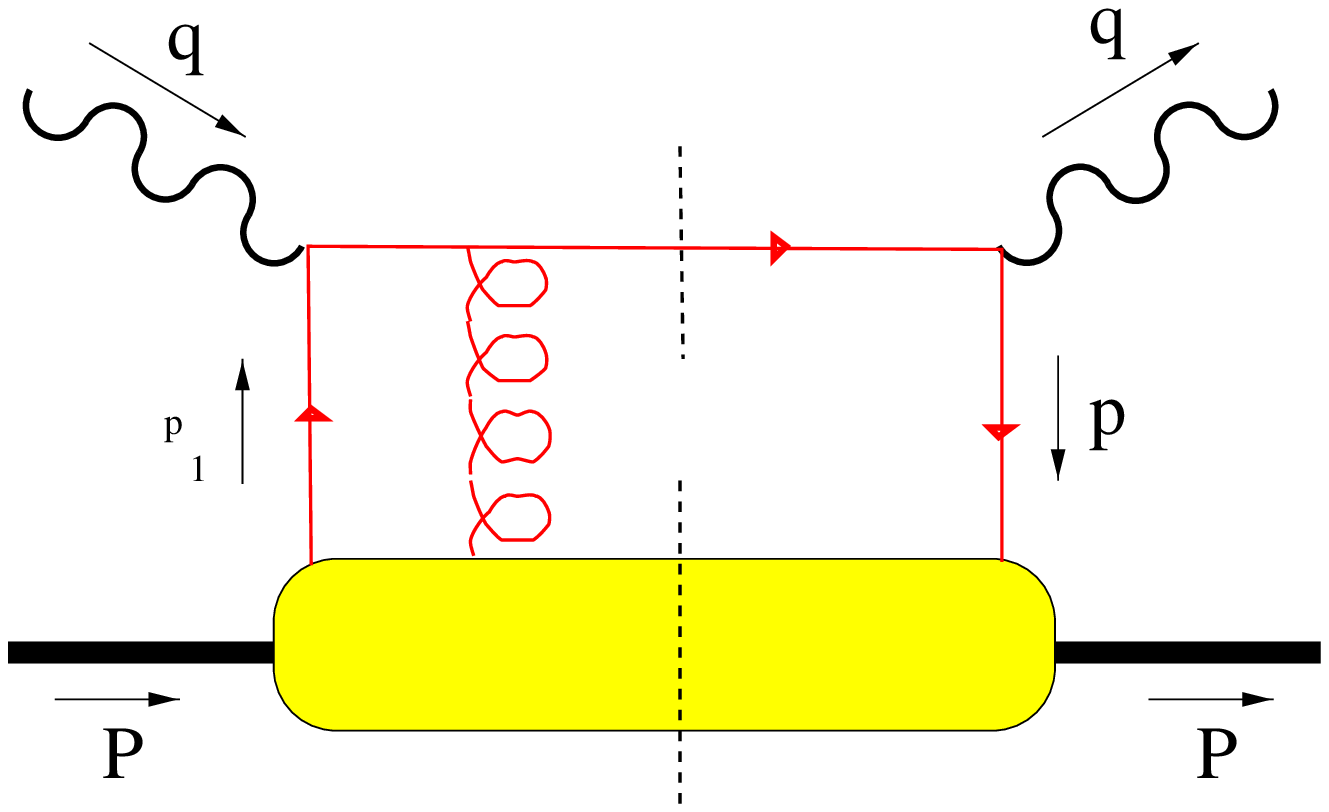,width=4.5cm}
\hspace{1cm}
\epsfig{file=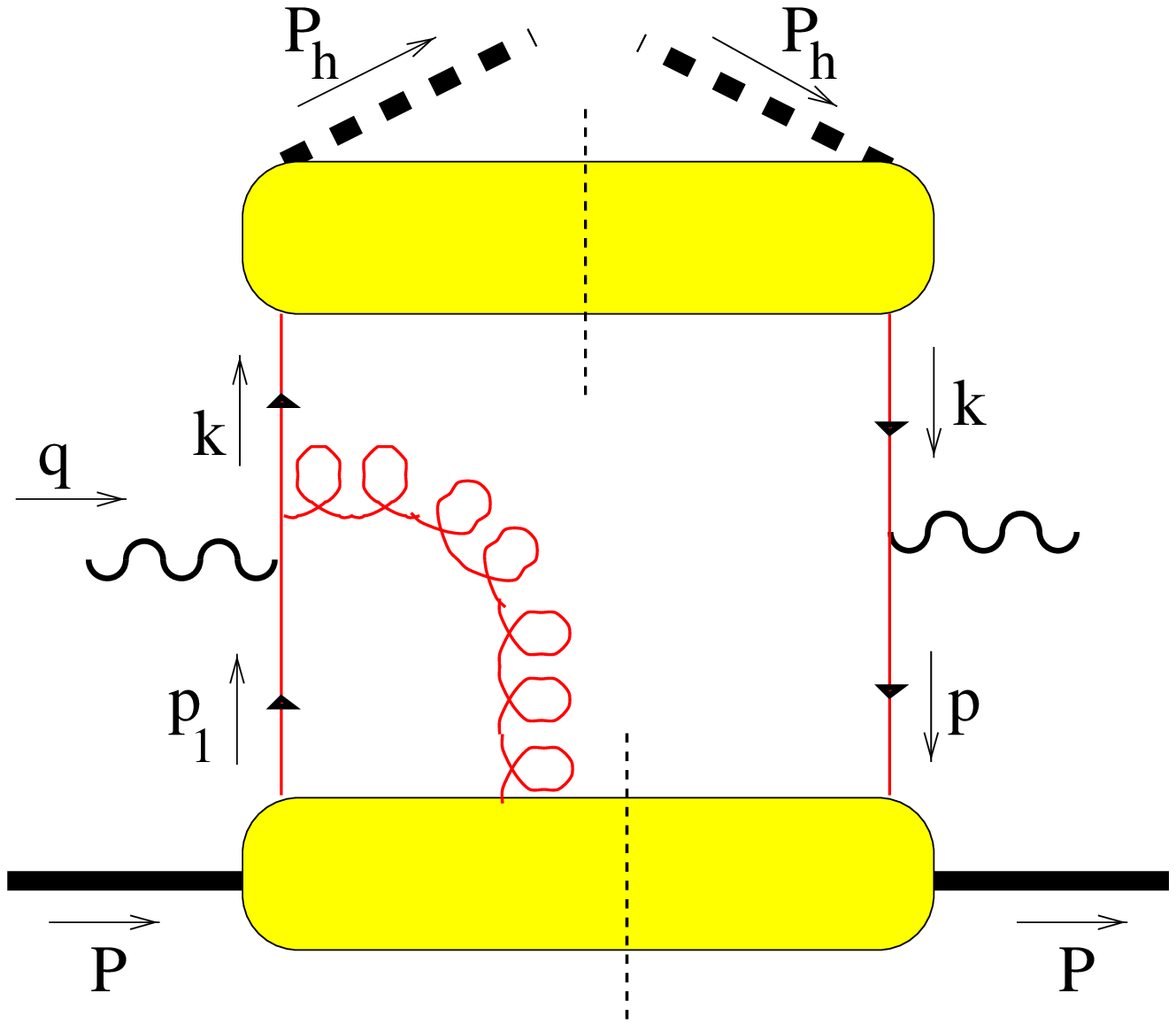,width=4.5cm}
\end{center}
\caption{\label{figdis1}
Examples of gluonic diagrams that must be included at subleading order
in lepton hadron inclusive scattering (left) and in semi-inclusive 
scattering (right).}
\end{figure}

There is a second contribution at order $1/Q$ coming from diagrams as the
one shown in Fig.~\ref{figdis1}. For these gluon diagrams one needs
matrix elements 
containing $\overline \psi(0)\,gA_\st^\alpha(\eta)\,\psi(\xi)$. At order
$1/Q$ one only needs the matrix element of the bilocal combinations
$\overline \psi(0)\,gA_\st^\alpha(\xi)\,\psi(\xi)$ and
$\overline \psi(0)\,gA_\st^\alpha(0)\,\psi(\xi)$
These soft parts have a structure quite similar to $\Phi_\partial^\alpha$
and are parametrized as
\bea
\Phi_A^\alpha (x) &= &
\frac{M}{2}\,\Biggl\{
-x\,\tilde g_{T}(x)\,S_\st^\alpha\,\slash n_+\gamma_5 
-\lambda\,x\,\tilde h_{L}(x)
\,\frac{[\gamma^\alpha,\slash n_+]\gamma_5}{2}
\nonumber \\
&&\quad \mbox{} 
-x\,\tilde f_T(x)
\,\epsilon^{\alpha}_{\ \ \mu\nu\rho}\gamma^\mu n_-^\nu {S_\st^\rho}
- x\,\tilde h(x)
\,\frac{i[\gamma^\alpha, \nslash_+]}{2}\Biggr\}.
\label{phiA}
\eea
This contributes also to $W_A^{\mu\nu}$,
\be
2M\,W_{A (b)}^{\mu\nu}({q,P,S_\st}) =
i\,\frac{2M\xbj}{Q}
\,\hat t_{\mbox{}}^{\,[\mu}\epsilon_\perp^{\nu ]\rho} {S_{\perp\rho}}
\,\tilde g_T(\xbj).
\ee
Using the QCD equations of motion, however, these functions can be related
to the functions appearing in $\Phi$. To be precise one combines $i\partial$ in
$\Phi_\partial$ (see Eq.~\ref{Phid})
and $A_\mu$ in $\Phi_A$ to $\Phi_D$ containing
$iD_\mu = i\partial_\mu + g\,A_\mu$ for which one has via the equations
of motion
\bea
\Phi_D^\alpha (x) &= &
\frac{M}{2}\,\Biggl\{
-\left(x\,g_{T}- \frac{m}{M}\,h_1\right)\,S_\st^\alpha\,\slash n_+\gamma_5 
\nonumber \\
&&\quad \mbox{} 
-\lambda\left(x\,h_{L}-\frac{m}{M}\,g_1\right)
\,\frac{[\gamma^\alpha,\slash n_+]\gamma_5}{2}
\nonumber \\
&&\quad \mbox{} 
-x\,f_T(x)
\,\epsilon^{\alpha}_{\ \ \mu\nu\rho}\gamma^\mu n_-^\nu {S_\st^\rho}
- x\,\tilde h(x)
\,\frac{i[\gamma^\alpha, \nslash_+]}{2}\Biggr\}.
\eea
Hence one obtains
\bea
&&
x\,\tilde g_T = x\,g_T - g_{1T}^{(1)} - \frac{m}{M}\,h_1,
\\
&&
x\,\tilde h_L = x\,h_L - h_{1L}^{\perp (1)} - \frac{m}{M}\,g_1,
\\
&&
x\,\tilde f_T = x\,f_T + f_{1T}^{(1)},
\\
&&
x\,\tilde h = x\,h + 2\,h_{1}^{\perp (1)}.
\eea
and one obtains the full contribution
\be
2M\,W_{A}^{\mu\nu}({q,P,S_\st}) =
i\,\frac{2M\xbj}{Q}
\,\hat t_{\mbox{}}^{\,[\mu}\epsilon_\perp^{\nu ]\rho} {S_{\perp\rho}}
\,g_T(\xbj),
\ee
leading for the structure function ${\it g}_T(\xbj,Q^2)$ defined in
Eq.~\ref{wanti} to the result
\be
{\it g}_T(\xbj,Q^2) = \frac{1}{2}\sum_a e_a^2
\left( g_T^a(\xbj) + g_T^{\bar a}(\xbj)\right).
\ee

{}From Lorentz invariance one obtains, furthermore, some interesting
relations between the subleading functions and the $k_\st$-dependent
leading functions~\cite{BKL84,MT96,BM98}. Just by using the expressions
for the functions in terms of the amplitudes $A_i$ in Eq.~\ref{lorentz}
one finds
\bea
&&g_T  = g_1 + \frac{d}{dx}\,g_{1T}^{(1)},
\label{gTrel}
\\
&&h_L = h_1 - \frac{d}{dx}\,h_{1L}^{\perp (1)},
\label{hLrel}
\\
&&f_T =  - \frac{d}{dx}\,f_{1T}^{\perp (1)},
\\
&&h =  - \frac{d}{dx}\,h_{1}^{\perp (1)}.
\label{rel4}
\eea
As an application, one can eliminate $g_{1T}^{(1)}$ using Eq.~\ref{gTrel}
and obtain (assuming sufficient neat behavior of the functions)
for $g_2 = g_T - g_1$
\bea
g_2(x) & = & -\left[ g_1(x) - \int_x^1 dy \,\frac{g_1(y)}{y} \right]
+ \frac{m}{M} \left[ \frac{h_1(x)}{x}-\int_x^1 dy\, \frac{h_1(y)}{y^2}\right]
\nonumber \\
&& \mbox{}
+ \left[\tilde g_T(x) - \int_x^1 dy\, \frac{\tilde g_T (y)}{y}\right].
\eea
One can use this to obtain for each quark flavor $\int dx\,g^a_2(x) = 0$, the
Burkhardt-Cottingham sumrule~\cite{BC}. 
Neglecting the interaction-dependent part one
obtains the Wandzura-Wilczek approximation~\cite{WW} 
for $g_2$, which in particular
when one neglects the quark mass term provides a simple and often used
estimate for $g_2$. It has become the standard with which experimentalists
compare the results for $g_2$. 

Actually the SLAC results for $g_2$ can also be used to estimate the
function $g_{1T}^{(1)}$ and the resulting asymmetries, e.g. the
one in Eq.~\ref{asbas}. For this one needs the exact relation
in Eq.~\ref{gTrel}. Results can be found in Refs~\cite{KM96} and
\cite{BM99}.

\subsection{Subleading 1-particle inclusive leptoproduction}

Also for the transverse momentum dependent functions dependent distribution 
and fragmentation functions one can proceed to subleading order
\cite{MT96}. We will not discuss these functions here. 

In semi-inclusive cross sections one also needs
fragmentation functions, for which similar relations exist, e.g. the
relation in Eq.~\ref{rel4} for distribution functions 
has an analog for fragmentation functions, relating
$H_1^{\perp (1)}$ (appearing in Eqs~\ref{as2} and \ref{finalstate})
and an at subleading order appearing function $H(z)$, 
\be
\frac{H(z)}{z} = z^2\,\frac{d}{dz} \left(\frac{H_1^{\perp (1)}}{z}\right).
\label{frag1}
\ee

An interesting subleading asymmetry involving $H_1^\perp$ is
a $\sin(\phi_h^\ell)$ single spin asymmetry appearing as the structure
functions ${\cal H}_{LT}^\prime$ in Eq.~{sidiswanti} for a polarized
lepton but unpolarized target~\cite{LM94},
\bea
&&\left< \frac{Q_{T}} {M} \,\sin(\phi_h^\ell) \right>_{LO} =
\nonumber \\ &&\qquad
\frac{4\pi \alpha^2\,s}{Q^4}\,{\lambda_e}
\,y\sqrt{1-y} 
\,\frac{2M}{Q}\,\xbj^2 {\tilde e^a}(\xbj)\,{H_1^{\perp (1)a}}(z_h)
\label{as1}
\eea
where $\tilde e^a(x) = e^a(x) - (m_a/M)\,(f_1^a(x)/x)$.
This cross section involves, besides the 
time-reversal odd fragmentation function $H_1^\perp$, 
the distribution function $e$.
The tilde function that appear in the cross sections is in fact
the socalled interaction dependent part of the twist three
functions. It would vanish in any naive parton model calculation in
which cross sections are obtained by folding electron-parton cross
sections with parton densities. Considering the relation for $\tilde e$
one can state it as $x\,e(x)$ = $(m/M)\,f_1(x)$ in the absence of
quark-quark-gluon correlations. The inclusion of the latter also
requires diagrams dressed with gluons as shown in Fig.~\ref{figdis1}.

\section{Color gauge invariance}

We have sofar neglected two problems. The first problem is that 
the correlation function $\Phi$ discussed in previous sections involve
two quark fields at different space-time points and hence are not
color gauge invariant. 
The second problem comes from the gluonic diagrams similar as the
ones we have discussed in the previous section (see Fig.~\ref{figdis1})
We note that diagrams involving matrix elements with longitudinal ($A^+$)
gluon fields,
\[
\overline \psi_j(0)\,gA^+(\eta)\,\psi_i(\xi)
\]
do not lead to any suppression. The reason is that because of the 
$+$-index in the gluon field 
the matrix element is proportional to $P^+$, $p^+$ or
$M\,S^+$ rather than the proportionality to $M\,S_\st^\alpha$ or
$p_\st^\alpha$ that we have seen in Eq.~\ref{phiA} for a gluonic
matrix element with transverse gluons.

A straightforward calculation, however, shows that the gluonic diagrams with
one or more longitudinal gluons involve matrix elements (soft parts)
of operators $\overline \psi \psi$, 
$\overline \psi\,A^+\,\psi$, $\overline \psi\,A^+A^+\,\psi$, etc.
that can be resummed into a correlation function
\be
\Phi_{ij}(x) =
\left. \int \frac{d\xi^-}{2\pi}\ e^{ip\cdot \xi}
\,\langle P,S\vert \overline \psi_j(0)\,{\cal U}(0,\xi)\,\psi_i(\xi)
\vert P,S\rangle \right|_{\xi^+ = \xi_\st = 0},
\ee
where ${\cal U}$ is a gauge link operator
\be
{\cal U}(0,\xi) 
= {\cal P}\exp\left(-i\int_0^{\xi^-} d\zeta^-\,A^+(\zeta)\right)
\ee
(path-ordered exponential with path along $-$-direction).
Et voila, the unsuppressed gluonic diagrams combine into 
a color gauge invariant correlation function.
We note that at the level of operators, one expands
\be
\overline \psi(0)\psi(\xi) = 
\sum_n \frac{\xi^{\mu_1}\ldots \xi^{\mu_n}}{n!}\,
\overline \psi(0)\partial_{\mu_1}\ldots\partial_{\mu_n}\psi(0),
\ee
in a set of local operators, but only
the expansion of the nonlocal combination with a gauge link 
\be
\overline \psi(0)\psi(\xi) = 
\sum_n \frac{\xi^{\mu_1}\ldots \xi^{\mu_n}}{n!}\,
\overline \psi(0)D_{\mu_1}\ldots D_{\mu_n}\psi(0),
\ee
is an expansion in terms of local gauge invariant operators. 
The latter operators are precisely the local (quark) operators
that appear in the operator product expansion applied to 
inclusive deep inelastic scattering.

For the $p_\st$-dependent functions, one finds that inclusion of
$A^+$ gluonic diagrams leads to a color gauge invariant matrix 
element with links running via $\xi^= = \pm \infty$. For instance
in lepton-hadron scattering one finds
\be
\Phi(x,\bm p_T) =
\left. \int \frac{d\xi^-d^2\bm \xi_T}{(2\pi)^3}\ e^{ip\cdot \xi}
\,\langle P,S\vert \overline \psi(0)\,{\cal U}(0,\infty)
\,{\cal U}(\infty,\xi)\,\psi(\xi)
\vert P,S\rangle \right|_{\xi^+ = 0},
\ee
where the gauge links are at constant $\xi_\st$.
One can multiply this correlator with $p_\st^\alpha$ and make this
into a derivative $\partial_\alpha$. Including the links one finds
the color gauge invariant result
\bea
&&p_\st^\alpha\,\Phi_{ij}(x,\bm p_\st) =
(\Phi_\partial^{\alpha})_{ij}(x,\bm p_\st) \nonumber \nonumber \\ &&
\quad =
\int \frac{d\xi^-\,d^2\bm \xi_\st}{(2\pi)^3}\ e^{ip\cdot \xi}
\biggl\{ \langle P,S\vert \overline \psi_j(0)
\,{\cal U}(0,\infty)\, iD_\st^\alpha\psi_i(\xi) \vert P,S\rangle
\biggr|_{\xi^+=0} 
\nonumber \\ && \mbox{}\hspace{2cm}
- \langle P,S\vert \overline \psi_j(0)\,{\cal U}(0,\infty)
\int_{\infty}^{\xi^-}d\eta^- \,{\cal U}(\infty,\eta)
\nonumber \\ && \mbox{}\hspace{3cm}\times g\,G^{+\alpha}(\eta)
\,{\cal U}(\eta,\xi)\,\psi_i(\xi) \vert P,S\rangle \biggr|_{\xi^+=0}\biggr\},
\eea
which gives after integration over $p_\st$ the expected result
$\Phi_\partial^\alpha(x) = \Phi_D^\alpha(x) - \Phi_A^\alpha(x)$. Note 
that in $A^+ = 0$ gauge all the gauge links disappear, while one
has $G^{+\alpha} = \partial^+A^\alpha$, but there presence is essential
to perform the above differentiations.

\section{Evolution}

\begin{figure}[t]
\begin{center}
\epsfig{file=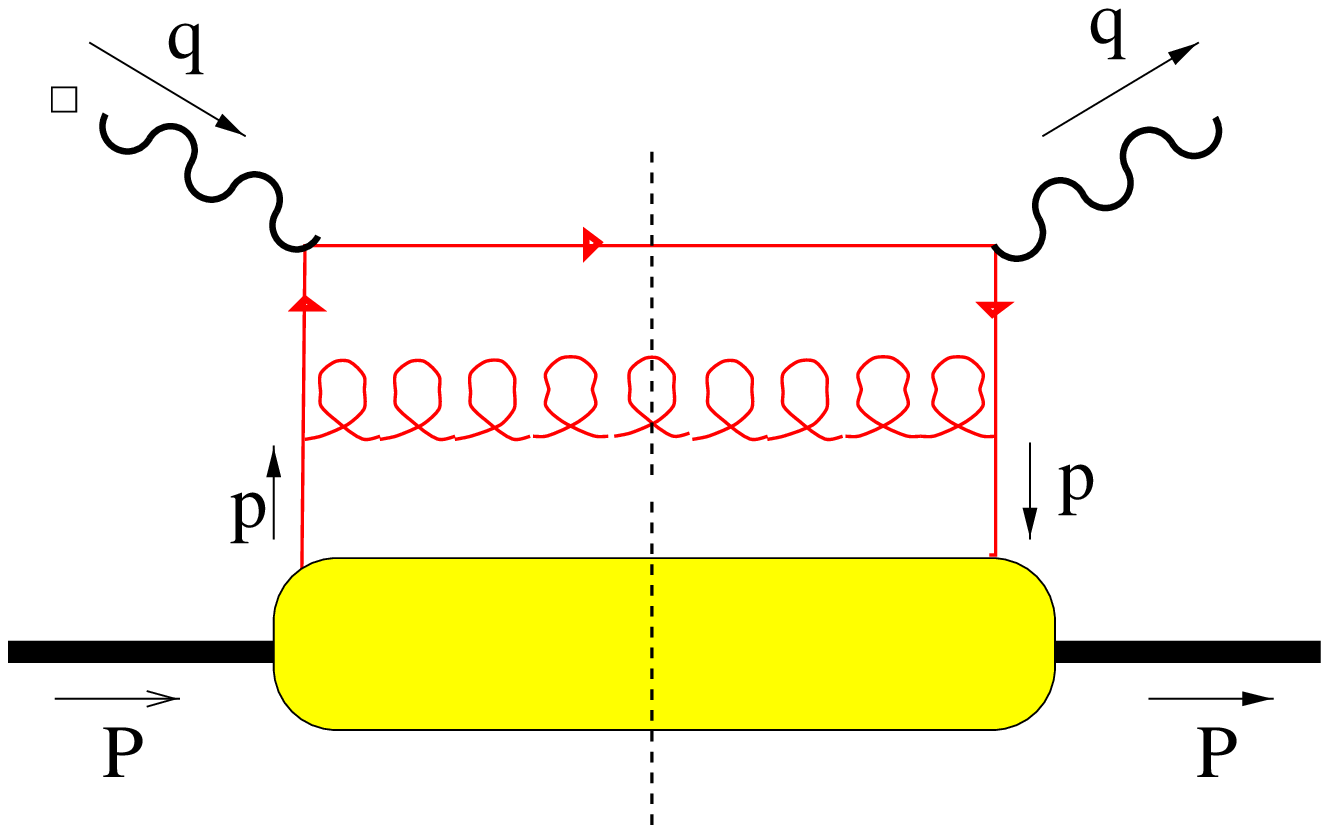,width=4.5cm}
\hspace{1cm}
\epsfig{file=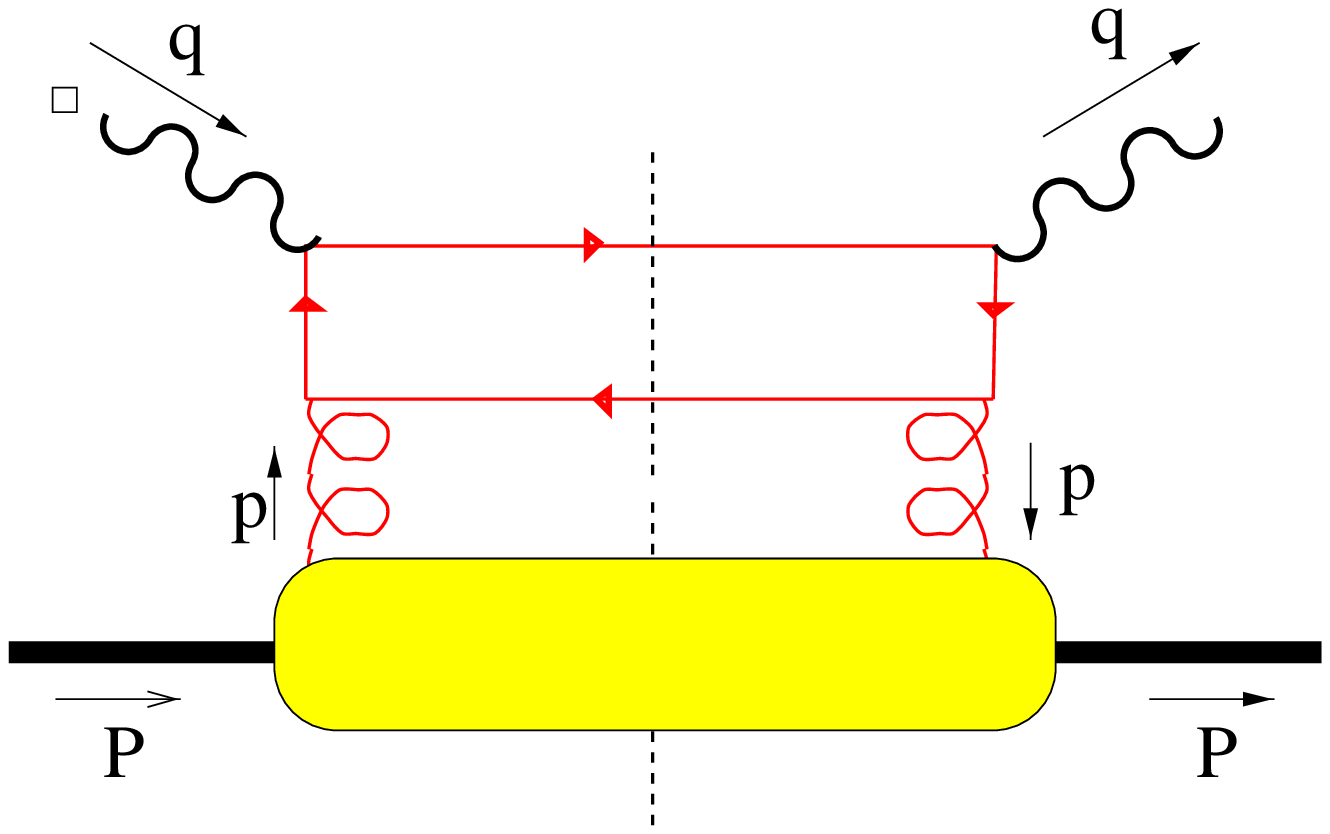,width=4.5cm}
\end{center}
\caption{\label{figdis2}
Ladder diagrams used to calculate the asymptotic behavior of the 
correlation functions.}
\end{figure}
The explicit treatment of transverse momenta provides also a transparent
way to include the evolution equations for quark distribution and 
fragmentation functions. Remember that we have assumed that soft parts
vanish sufficiently fast as a function of the invariants $p\cdot P$ and
$p^2$, which at constant $x$ implies a sufficiently fast vanishing as
a function of $\bm p_\st^2$. This simply turns out not to be true.
Assuming that the result for $\bm p_\st^2 \ge \mu^2$ is given by
the diagram shown in Fig.~\ref{figdis2} one finds
\bea
f_1(x,\bm p_\st^2) & = & \theta (\mu^2-\bm p_\st^2)\,f_1(x,\bm p_\st^2)
\nonumber \\ & + &
\theta(\bm p_\st^2-\mu^2)
\,\frac{1}{\pi\,\bm p_\st^2}
\,\frac{\alpha_s(\mu^2)}{2\pi}
\int_x^1 \frac{dy}{y}\,P_{qq}\left(\frac{x}{y}\right)\,f_1(y;\mu^2),
\eea
where $f_1(x;\mu^2) = \pi \int_0^{\mu^2} d\bm p_\st^2\ f_1(x,\bm p_\st^2)$
and the splitting function is given by
\be
P_{qq}(z) = C_F\,\left[\frac{1+z^2}{(1-z)_+} + \frac{3}{2}\,\delta(1-z)\right],
\ee
with $\int dz\,f(z)/(1-z)_+ \equiv \int dz\,(f(z)-f(1))/(1-z)$
and the color factor $C_F$ = 4/3 for $SU(3)$.
With the introduction of the scale in $f_1(x;\mu^2)$ one sees that the
scale dependence satisfies
\be
\frac{\partial f_1(x;\mu^2)}{\partial \ln \mu^2}
= \frac{\alpha_s}{2\pi}
\int_x^1 \frac{dy}{y}\,P_{qq}\left(\frac{x}{y}\right)\,f_1(y;\mu^2).
\ee
This is the standard~\cite{Roberts}
nonsinglet evolution equation for the valence quark distribution
function. For the flavor singlet combination of quark distributions or the
sea distributions one also needs to take into account contributions as
shown in Fig.~\ref{figdis2} (right) involving the gluon distribution
functions related to matrix elements with gluon fields $F_{\mu\nu}(\xi)$
but otherwise proceeding along analogous lines.
The $\delta$-function contribution can be explicitly calculated by including
vertex corrections (socalled virtual diagrams), but it is easier to derive
them by requiring that the sum rules for $f_1$ remain valid under evolution,
which requires that $\int_0^1 dz\ P_{qq}(z) = 0$.
\begin{quotation}
\small
Except for logarithmic contributions also finite $\alpha_s$ contributions 
show up in deep inelastic scattering~\cite{Roberts}. 
For instance in inclusive scattering
one finds that the lowest order result for $F_L$ is of this type, 
\begin{eqnarray}
F_L(\xbj,Q^2) & = & \frac{\alpha_s(Q^2)}{4\pi} \Biggl[ C_F \int_{\xbj}^1
\frac{dy}{y}\,\left(\frac{2\xbj}{y}\right)^2 y\,f_1(y;Q^2) \nonumber \\
&&\mbox{} + \left( 2\sum_q e_q^2 \right) \int_{\xbj}^1
\frac{dy}{y} \,\left( \frac{2\xbj}{y} \right)^2 \left( 1 - \frac{\xbj}{y}
\right) \,y\,G(y;Q^2) \Biggr],
\nonumber \\ &&
\end{eqnarray}
the second term involving the gluon distribution function $G(x)$.
\end{quotation}

\section{Concluding remarks}

In these lectures I have discussed aspects of hard scattering processes,
in particular inclusive and 1-particle inclusive lepton-hadron scattering.
The goal is the study of the quark and gluon structure of hadrons. 
For example, by considering polarized targets or particle production 
one can measure spin and azimuthal asymmetries and use them to obtain
information on specific correlations between
spin and momenta of the partons. The reason why this is a promising 
route is the existence of a field theoretical framework that allows
a clean study of the observables as well-defined hadronic matrix elements. 

\section*{Acknowledgements}

I acknowledge many discussions with colleagues, in particular Elliot Leader,
Stan Brodsky.


\begin{thebibliography}{99}
\bibitem{Roberts}
R.G. Roberts, {\em The structure of the proton}, 
Cambridge University Press 1990
\bibitem{RS79}
J.P. Ralston and D.E. Soper, 
Nucl. Phys. {\bf B152 (1979)} 109.
\bibitem{TM95}
R.D. Tangerman and P.J. Mulders, Phys. Rev. {\bf D51} (1995) 3357.
\bibitem{Soper77}
D.E. Soper, 
Phys. Rev. {\bf D 15} (1977) 1141; 
Phys. Rev. Lett.  43 (1979) 1847.
\bibitem{Jaffe83}
R.L. Jaffe, 
Nucl. Phys. {\bf B 229} (1983) 205.
\bibitem{JJ92}
R.L. Jaffe and X. Ji, 
Nucl. Phys. {\bf B 375} (1992) 527.
\bibitem{Artru}
X. Artru and M. Mekhfi, Z. Phys. {\bf 45} (1990) 669,
J.L. Cortes, B. Pire and J.P. Ralston, Z. phys. {\bf C55} (1992) 409.
\bibitem{Soffer}
J. Soffer and D. Wray, Phys. Lett. {\bf 43B} (1973) 514.
\bibitem{bmt}
N. Hammon, O. Teryaev and A. Sch\"afer, 
Phys. Lett. {\bf B390} (1997) 409;
D. Boer, P.J. Mulders and O.V. Teryaev, 
\bibitem{Sivers90}
D. Sivers,
Phys. Rev. {\bf D41} (1990) 83 and
Phys. Rev. {\bf D43} (1991) 261. 
\bibitem{Anselmino95}
M. Anselmino, M. Boglione and F. Murgia,
Phys. Lett. {\bf B362} (1995) 164;
M. Anselmino and F. Murgia, Phys. Lett. {\bf B442} (1998) 470.
\bibitem{CS82}
J.C. Collins and D.E. Soper, 
Nucl. Phys. {\bf B 194} (1982) 445.
\bibitem{Collins93}
J. Collins, Nucl. Phys. {\bf B396} (1993) 161.
\bibitem{HHK83}
K. Hagiwara, K. Hikasa and N. Kai, 
Phys. Rev. {\bf D27} (1983) 84.
\bibitem{JJ93}
R.L. Jaffe and X. Ji, Phys. Rev. Lett. {\bf 71} (1993) 2547.
Phys. Rev. {\bf D57} (1998) 3057.
\bibitem{LM94}
J. Levelt and P.J. Mulders, Phys. Rev. {\bf D 49} (1994) 96;
Phys. Lett. {\bf B 338} (1994) 357.
\bibitem{Efremov}
A.V. Efremov, O.G. Smirnova, L.G. Tkachev, in proceedings of the
13. International Symposium on High-Energy Spin Physics (SPIN98), 
Protvino, Russia, 8-12 Sep 1998, Nucl. Phys. Suppl. {\bf 74} (1999) 49.
\bibitem{KM96}
A. Kotzinian and P.J. Mulders, Phys. Rev. {\bf D54} (1996) 1229.
\bibitem{Kotzinian95} 
A. Kotzinian, Nucl. Phys. {\bf B 441} (1995) 234.
\bibitem{TM95b}
R.D. Tangerman and P.J. Mulders, Phys. Lett. {\bf B352} (1995) 129.
\bibitem{BKL84}
A.P. Bukhvostov, E.A. Kuraev and L.N. Lipatov, 
Sov. Phys. {\bf JETP 60} (1984) 22.
\bibitem{MT96}
P.J. Mulders and R.D. Tangerman,
Nucl. Phys. {\bf B461} (1996) 197.
\bibitem{BM98}
D. Boer and P.J. Mulders, Phys. Rev. {\bf D 57} (1998) 5780.
\bibitem{BC}
H. Burkhardt and W.N. Cottingham, Ann. Phys. (N.Y.) {\bf 56} (1976) 453.
\bibitem{WW} 
S. Wandzura and F. Wilczek, Phys. Rev. {\bf D16} (1977) 707.
\bibitem{BM99}
M. Boglione and P.J. Mulders, 
Phys. Rev. {\bf D60} (1999) 054007.
\end{thebibliography}
\end{document}